\documentclass[%
aps,prb
,reprint
,superscriptaddress%
,showpacs
,amssymb,amsmath,amsfonts
]{revtex4-1}
\usepackage{docs}%
\usepackage{graphicx}%
\usepackage{comment}%
\usepackage[colorlinks=true,linkcolor=blue]{hyperref}%
\begin{document}
\title{Multi-Stability and Condensation of Exciton-Polaritons below Threshold}

\author{Jiun-Yi Lien}%
\email{JiunYi.Lien@gmail.com}%
\affiliation{Department of Physics and Center for Theoretical Sciences, National Cheng Kung University, Tainan 701, Taiwan}
\affiliation{Department of Engineering Science, National Cheng Kung University, Tainan 701, Taiwan}%
\author{Yueh-Nan Chen}%
\email{yuehnan@mail.ncku.edu.tw}%
\affiliation{Department of Physics and Center for Theoretical Sciences, National Cheng Kung University, Tainan 701, Taiwan}
\author{Natsuko Ishida}%
\affiliation{CEMS, RIKEN, Wako-shi, Saitama 351-0198, Japan}
\author{Hong-Bin Chen}
\affiliation{Department of Physics and Center for Theoretical Sciences, National Cheng Kung University, Tainan 701, Taiwan}
\author{Chi-Chuan Hwang}%
\affiliation{Department of Engineering Science, National Cheng Kung University, Tainan 701, Taiwan}%
\author{Franco Nori}%
\affiliation{CEMS, RIKEN, Wako-shi, Saitama 351-0198, Japan}
\affiliation{Physics Department, University of Michigan, Ann Arbor, MI 48104-4313, USA}

\date{\today}%
\begin{abstract}
Exciton-polaritons can condense to a macroscopic quantum state through a non-equilibrium process of pumping and decay.
In recent experiments, polariton condensates are used to observe, for a short time, nonlinear Josephson phenomena by coupling two condensates.
However, it is still not clear how these phenomena are affected by the pumping and decay at long times and how the coupling alters the polariton condensation.
Here, we consider a polariton Josephson junction pumped on one side and study its dynamics within a mean-field theory. 
The Josephson current is found to give rise to multi-stability of the stationary states, which are sensitive to the initial conditions and incoherent noises.
These states can be attributed to either the self-trapping effect or the parity-time (PT) symmetry of the system.
These results can be used to explain the emission spectra and the $\pi$-phase locking observed in recent experiments.
We further predict that the multi-stability can reduce to the self-trapped state if the PT symmetry is broken.
Moreover, the polaritons can condense even below the threshold, exhibiting hysteresis.
\end{abstract}


\maketitle%
\section{Introduction}

Exciton-polaritons, i.e., quasi-particles composed of cavity photons and quantum-well excitons in semiconducting microcavities, have recently been demonstrated to form a Bose-Einstein condensate (BEC) due to their light effective mass originating from the photonic part.
\cite{Kasprzak:Nature(2006)(BEC_exciton_polaritons),Balili:Science(2007)(Polariton_BEC_trap),Carusotto&Ciuti:RevModPhys.85.299(2013)(Quantum_fluids_light),Yamaguchi&Yamamoto:PRL.111.026404(2013)(2nd_Thresholds_BEC-BCS-Laser),Schneider&Yamamoto:Nature.497.348(2013)(electrically_pumped_polariton_laser)} 
The exciton-polariton BEC can be used to investigate macroscopic quantum effects in semiconducting systems, 
such as superfluidity
\cite{Amo:Nature.457.291(2009)(Polariton_BEC_superfluid),Tosi&Baumberg:NPhy(8)190(2012)(Sculpting_oscillators_polariton_BEC)} 
and quantized vortices.
\cite{Yamamoto:NPhy(7)129(2011)(Polariton_BEC_vortex-antivortex),Lagoudakis&Wouters:NPhy.4.706(2008)(Polariton_BEC_vortiex)} 
In addition, the well-known Josephson effects can be exhibited in a bosonic Josephson junction (BJJ), i.e., two coherently coupled condensates confined in a double-well potential.
Besides the phase-difference-induced current (d.c.~effect) and the detuning-induced oscillations (a.c.~effect) observed in superconducting and helium superfluid systems,
the interaction between the condensate bosons leads to nonlinear effects when the boson number increases, 
such as anharmonic oscillations and initial-imbalance-induced trapping toward one site (called macroscopic quantum self-trapping).
\cite{Raghavan:PRA.59.620(1999)(BEC_Josephson_effects)}
These phenomena were first observed in $^{87}\textrm{Rb}$ condensates,
\cite{Albiez:PRL.95.010402(2005)(Atomic_BEC_self-trapping)} 
and have also been demonstrated recently in exciton-polariton BJJs with disordered double wells 
\cite{Lagoudakis&Wouters:PRL.105.120403(2010)(Polariton_BEC_Josephson_Junction)} 
and semiconductor micropillars. 
\cite{Abbarchi:NPhy(2013)(Polariton_BEC_Josephson_oscillations)}

Since the lifetime of exciton-polaritons is about tens to hundreds of picoseconds, an optical pumping is necessary to compensate both the radiative and non-radiative decay of the polaritons. 
Therefore, the condensates are formed in such a non-equilibrium process. 
At low temperatures, one can use a non-resonant pumping to excite higher-energy excitons, which rapidly relax to form an incoherent polariton reservoir at the bottleneck of the lower-energy polariton band through the exciton-phonon interactions. 
The polaritons are then stimulatedly scattered (or cooled) into the ground state to achieve condensation with spontaneous coherence, provided that the incoherent polariton density reaches the threshold density. 
\cite{Kasprzak:Nature(2006)(BEC_exciton_polaritons),Porras&Ciuti:PRB.66.085304(2002)(Polariton_BEC_dynamics)}
The polariton reservoir drastically changes the Bogoliubov dispersion of the elementary excitations for condensates with conserved particle numbers, leading to a diffusive behavior in the long wavelength regime and the dynamical instability for the reservoir lifetime being comparable to the condensate. 
\cite{Szymanska&Keeling&Littlewood:PRL.96.230602(2006)(Polariton_BEC_incoherently_pumped_dynamics),Wouters:PRL(99)140402(2007)(polariton_BEC_excitations)}

While the dynamics of the polariton BJJ for different regimes can be accessed by a pulsed resonant excitation that gives the appropriate initial conditions,
\cite{Abbarchi:NPhy(2013)(Polariton_BEC_Josephson_oscillations)} 
the short-time behavior of the polariton BJJ is also affected by the reservoir, e.g., the spontaneous coherent oscillations driven by a non-resonant pumping covering the double well.
\cite{Lagoudakis&Wouters:PRL.105.120403(2010)(Polariton_BEC_Josephson_Junction)} 
In addition to these short-time oscillations, the non-equilibrium process with a continuous-wave (cw) pumping laser would also alter the behavior at long times.
It has been studied 
\cite{Wouters:PRB.77.121302(2008)(Synchronization_phase_transition_polariton_BJJ)} 
that under equivalent pumping on both sides, the synchronization of the condensates with a single eigen-energy can be destroyed by the potential difference across the junction, 
where a long-time a.c.~Josephson oscillation exists in the desynchronized phase.
\cite{Borgh:PRB.81.235302(2010)(Spinor_polariton_BEC)} 
A similar synchronization-desyncronization phase transition was also observed \cite{Baas&Lagoudakis:PRL.100.170401(2008)(Synchronization_phase_transition_polariton_BEC_disorder)} in microcavities with in-plane disorder.

\begin{figure*}[t]
  \includegraphics[width=0.35\linewidth]{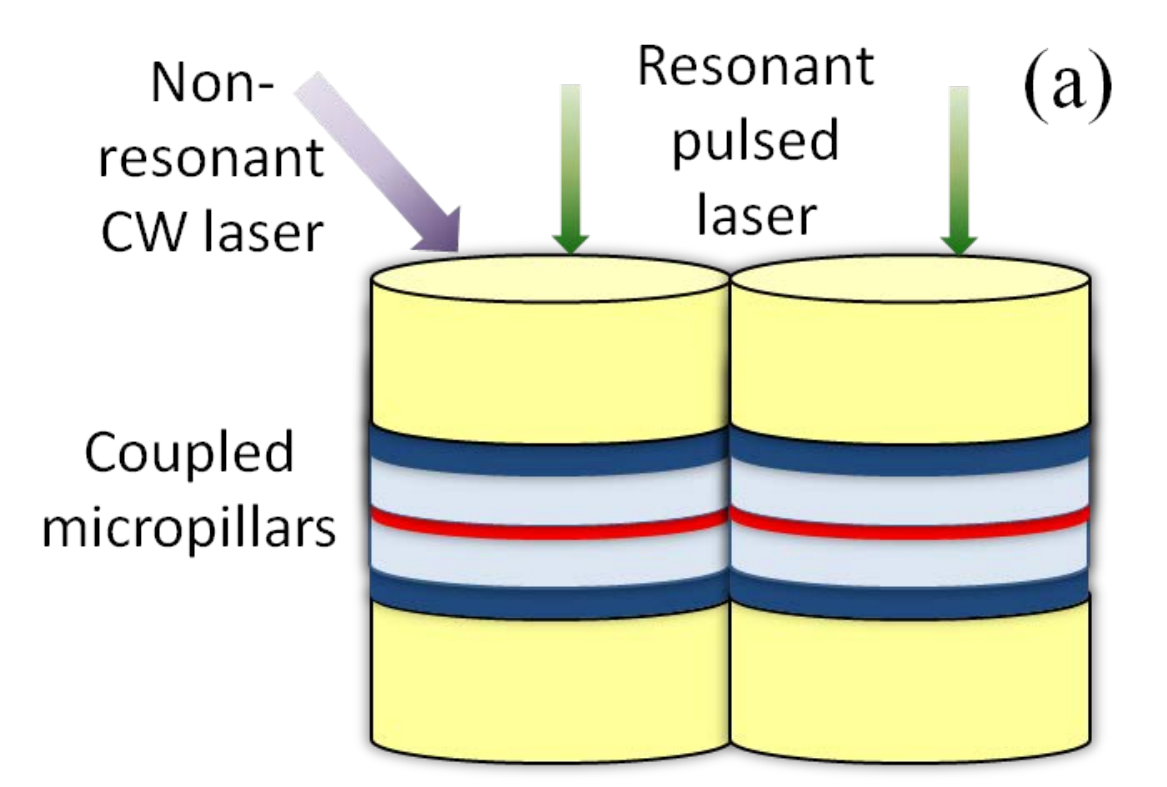}
  \qquad
  \includegraphics[width=0.6\linewidth]{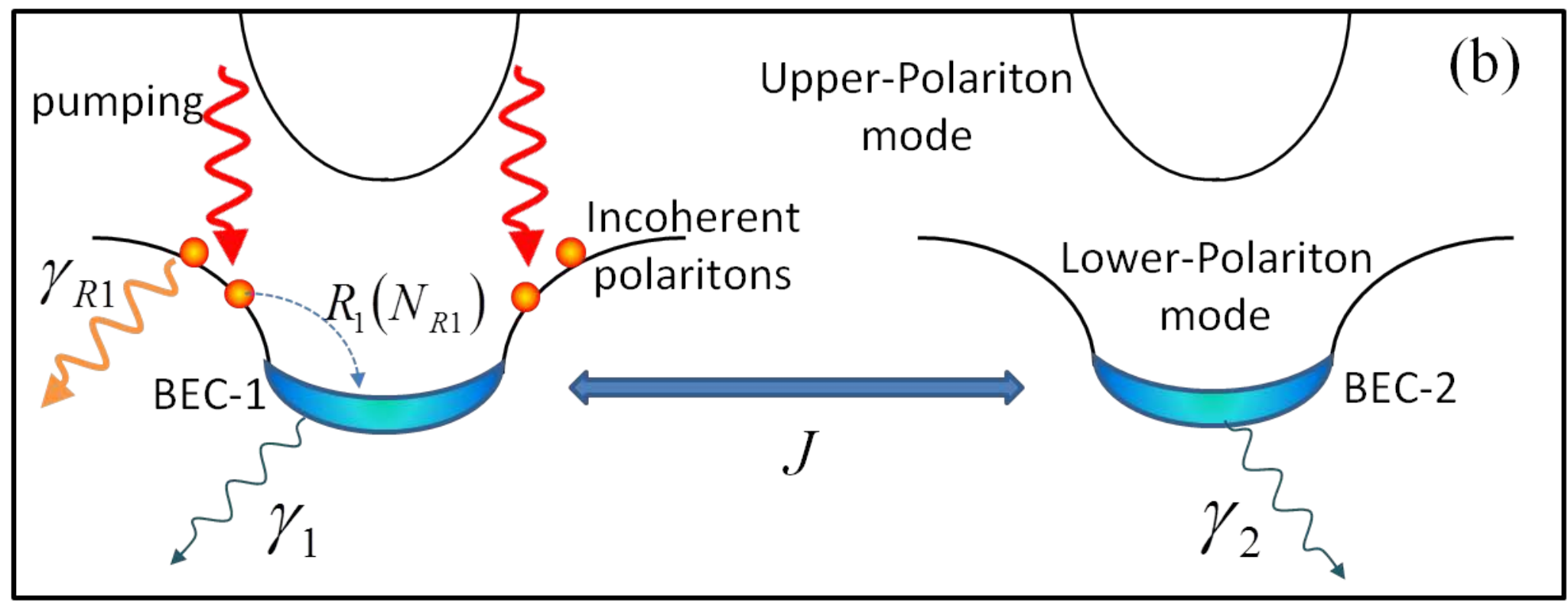}
  \caption{(Color online)
  (a) a polariton Josephson junction made of semiconducting micropillars pumped on one side.
  (b) the model of exciton-polariton condensation of the junction.
  The non-resonant continuous-wave laser is used to excite high-energy excitons as a pumping to the polariton reservoir, and the resonant pulsed laser can be used to control the initial condition.
  }\label{Transverse_coupled_polariton_BEC}
\end{figure*}

In this work, we consider a different situation: a polariton BJJ with a single non-resonant pumping focused on one side [Fig.~\ref{Transverse_coupled_polariton_BEC}(a)]. 
This has been realized in recent experiments using a pumping laser with small enough spot size.
\cite{Abbarchi:NPhy(2013)(Polariton_BEC_Josephson_oscillations),Galbiati:PRL.108.126403(2012)(Polariton_BEC_Photonic_Molecules)} 
In contrast with the regime under equivalent pumping,
\cite{Wouters:PRB.77.121302(2008)(Synchronization_phase_transition_polariton_BJJ),Borgh:PRB.81.235302(2010)(Spinor_polariton_BEC)} 
no a.c.~Josephson oscillations are found even in the presence of potential difference.
We will present the multiple stability induced by the Josephson current, as well as its effect on the threshold pumping.
Furthermore, we also find the possibility of condensation even below the threshold.

\section{Generalized Gross-Pitaevskii equations}

Our analysis is based on the generalized Gross-Pitaevskii equation (GPE) of the BJJ wave functions $\vec{\Psi}\equiv(\Psi_{1},\Psi_{2})^{T}\equiv(\sqrt{N_{c1}}e^{i\varphi_{1}},\sqrt{N_{c2}}e^{i\varphi_{2}})^{T}$, 
\begin{align}
  i{d\over dt}
  \vec{\Psi}
  &=
  H
  \vec{\Psi}
  \,,\label{GGPE}
\end{align}
coupled to a rate equation describing the reservoir.
\cite{Wouters:PRL(99)140402(2007)(polariton_BEC_excitations),Wouters:PRB.77.121302(2008)(Synchronization_phase_transition_polariton_BJJ)} 
The nonlinear and non-Hermitian Hamiltonian is written as
\begin{gather}
  H
  \equiv
  \left(
  \begin{array}{cc}
  E_{1} & -J \\
  -J & E_{2}
  \end{array}
  \right)
  \,,
  \nonumber\\
  E_{j}
  \equiv
  \epsilon_{j}
  +
  V_{j}\left(N_{Rj}\right)
  +
  U_{j}
  |\Psi_{j}|^2
  \,,\label{On-site energy}
\end{gather}
where $\epsilon_{j}$ is the single-particle ground state energy, $U_{j}$ is the condensate charging energy, and $J$ is the tunneling between the two sites.
Here the local dispersion is ignored and we only consider the ground-state wave function of each site, as shown in 
Fig.~\ref{Transverse_coupled_polariton_BEC}(b).

The complex effective potential $V_{j}$ is given by
\begin{align}
  V_{j}\left(N_{Rj}\right)
  &\equiv
  \left(
  \frac{\tilde{g}}{A_{j}}
  N_{Rj}
  +
  \mathcal{G}_{j}
  P_{j}
  \right)
  +
  \frac{i}{2}
  \left[
  R_{j}\left(N_{Rj}\right)
  -
  \gamma_{j}
  \right]
  \nonumber\\
  &\equiv
  V_{j}^{R}\left(N_{Rj}\right)
  +
  i
  V_{j}^{I}\left(N_{Rj}\right)
  \,,
\end{align}
and it depends on the number of reservoir polaritons $N_{Rj}(t)$ and the pumping strength $P_{j}$,
where $\tilde{g}$ is the interaction between the condensate and reservoir polaritons, $A_{j}$ is approximately the distribution area of the reservoir, 
$\mathcal{G}_{j}$ corresponds to the interaction between the condensate and high-energy excitons, 
\cite{Wouters&Carusotto&Ciuti:PRB.77.115340(2008)(Polariton_BEC_spatial&spectral_shape)}
and $\gamma_{j}$ is the decay of the condensates.
The term $R_{j}\left(N_{Rj}\right)$ is the stimulated scattering from the reservoir to the condensate, and, for simplicity, we only consider it as a linear function, i.e.
$
  R_{j}\left(N_{Rj}\right)
  \equiv
  R_{j}'N_{Rj}
$.
The rate equation of the reservoir on the pumped site is given by
\begin{align}
  {d\over dt}
  N_{R1}
  &=
  P_{1}
  -
  \gamma_{R1}
  N_{R1}
  -
  R_{1}(N_{R1})
  |\Psi_{1}|^2
  \,,\label{reservoir rate eq.}
\end{align}
where the reservoir decay $\gamma_{R1}$ and scattering loss are balanced by the laser pumping $P_{1}$. 
We can ignore $N_{R2}$ due to the weak diffusion of the reservoir polaritons.
\cite{Wouters:PRL(99)140402(2007)(polariton_BEC_excitations)} 
The interaction terms in $V_{j}^{R}$ can be treated as the effective potential energies together with $\epsilon_{j}$.
We ignore $\epsilon_{j}$ and $V_{j}^{R}$ in Sec.~\ref{stationary states without detuning}, and the effects of these potential differences (detuning) will be discussed later in \ref{Detuning}.


\section{Stationary states}\label{stationary states without detuning}

\subsection{Non-condensed solutions}

\begin{figure}[t]
  \includegraphics[width=\linewidth]{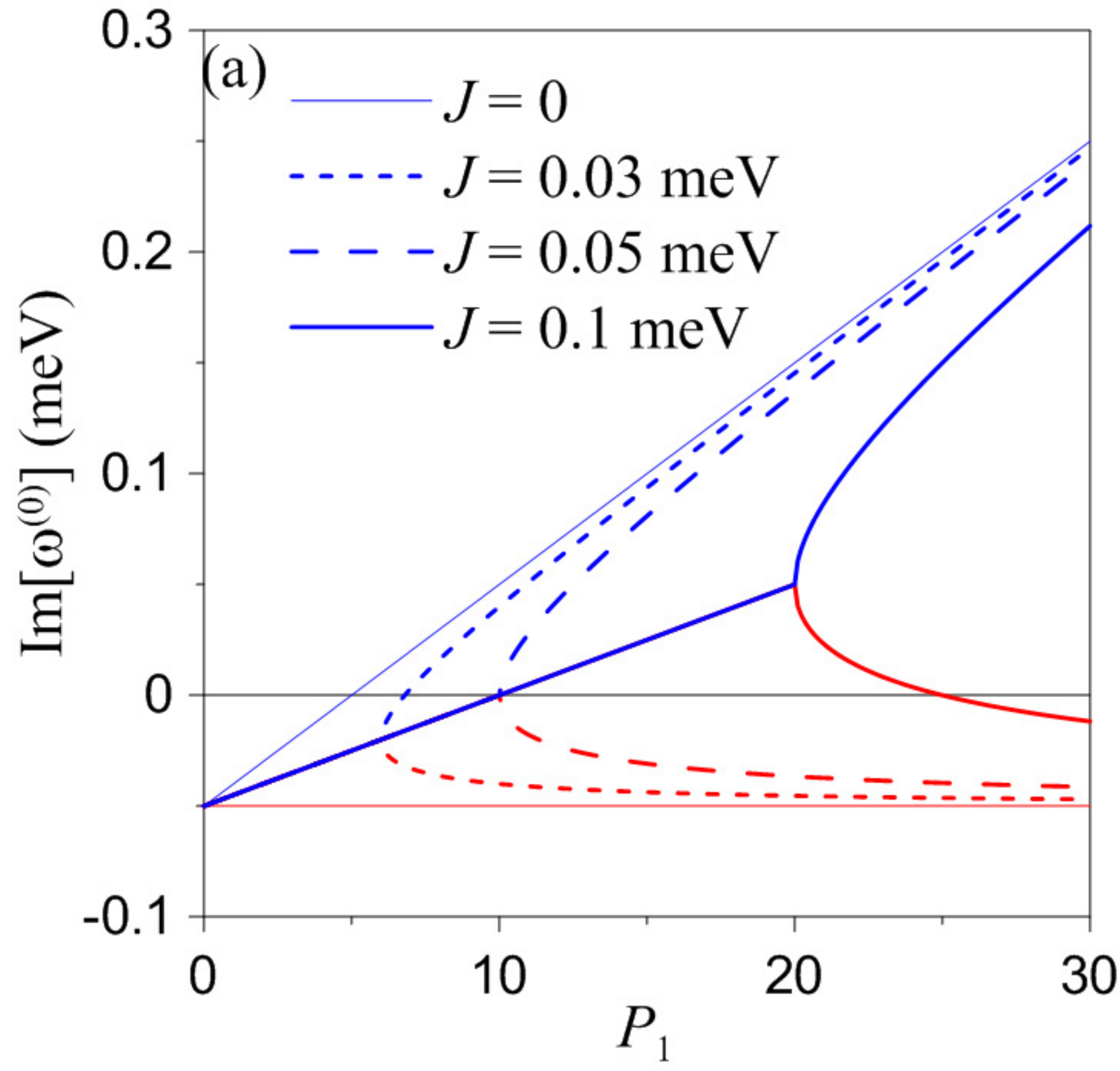}
  \\
  \includegraphics[width=\linewidth]{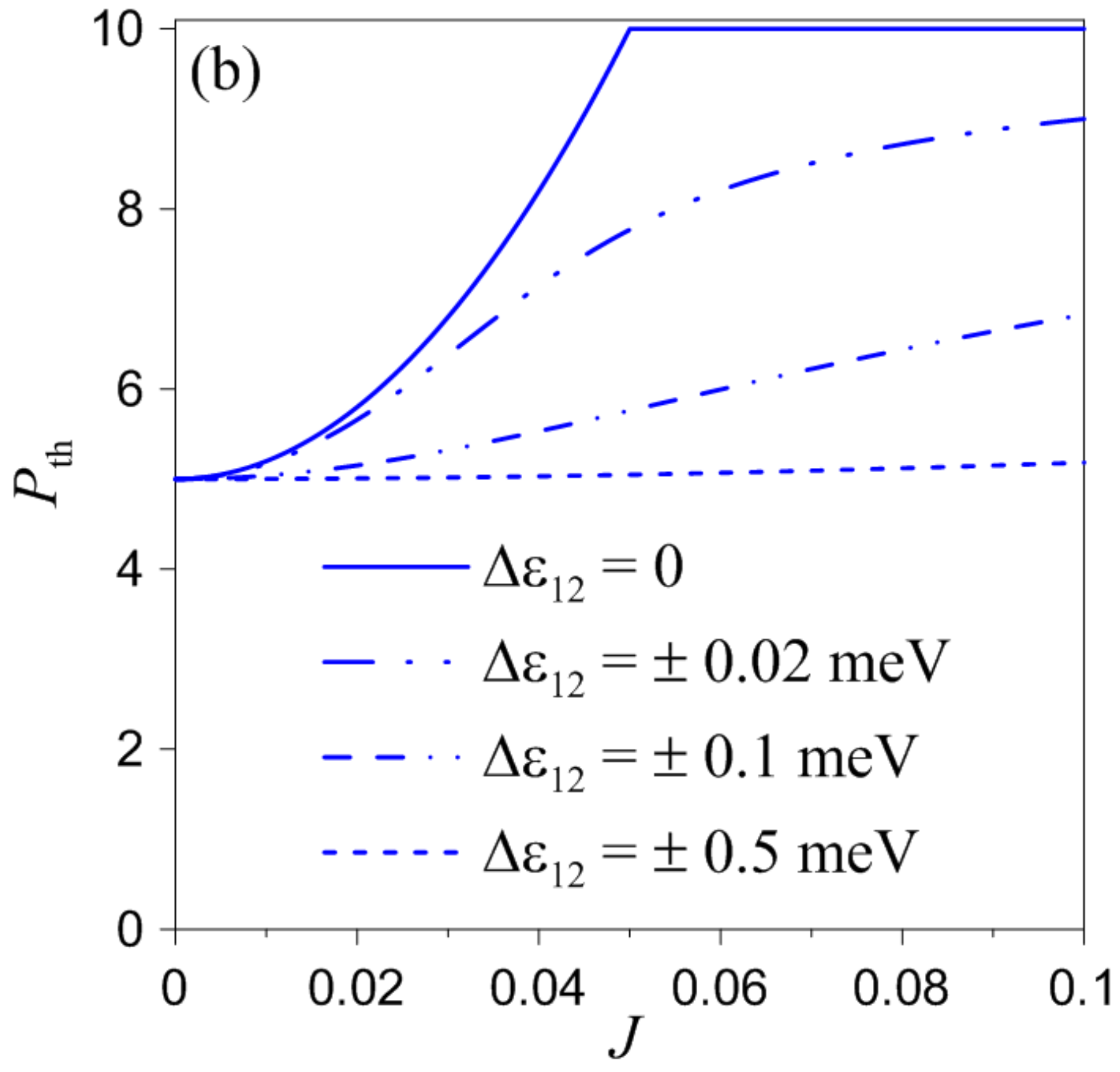}
  \caption{(Color online)
  (a) The imaginary part of the excitation spectrum $\omega^{(0)}$ without condensation.
  (b) Threshold pumping $P_{\textnormal{th}}$ as a function of the Josephson coupling strength $J$ with and without the detuning $\Delta\varepsilon_{12}$, where $P_{\textnormal{th}}$ is defined as the first point of $\textnormal{Im}[\omega_{\pm}^{(0)}]=0$ in (a). The decay rates are $\gamma_{1}=\gamma_{2}=0.1$ meV, which are close to the micropillar experiment, $\gamma_{R1}=0.5$ meV, and $R_{1}'=0.01$ meV.
  }\label{Threshold}
\end{figure}

A trivial solution is that no polaritons are condensed and the reservoir-polariton number is proportional to the pumping from (\ref{reservoir rate eq.}). 
By linearizing Eq.~(\ref{GGPE}), the fluctuation spectrum can be derived as
\begin{align}
  \omega_{\pm}^{(0)}
  &=
  \frac{1}{2}
  \left(
  E_{1}^{(0)}
  +
  E_{2}^{(0)}
  \right)
  \pm
  \frac{1}{2}
  \sqrt{
  \left(
  E_{1}^{(0)}
  -
  E_{2}^{(0)}
  \right)^2
  +
  4J^2
  }
  \,,\label{non-condensation excitation energy}
\end{align}
where $E_{j}^{(0)}$ is given by Eq.~(\ref{On-site energy}) with $\vec{\Psi}=0$.
As shown in Fig.~\ref{Threshold}, the threshold pumping $P_{\textnormal{th}}=\gamma_{R1}N_{R\textnormal{th}}$ can be determined by the first point of $\textnormal{Im}[\omega_{\pm}^{(0)}]=0$, where the non-condensed solution becomes unstable.
For $J=0$, the threshold reduces to the single-BEC case with $R_{1}(N_{R\textnormal{th}})=\gamma_{1}$.
It increases with $J$ until $N_{R\textnormal{th}}$ reaches a saturation point with $R_{1}(N_{R\textnormal{th}})=\gamma_{1}+\gamma_{2}$. 

\subsection{Real spectrum of the condensates}

In general, the spectrum of the non-Hermitian Hamiltonian $H$ is a complex function of the polariton numbers of the condensates and the reservoir.
To solve the nonzero stationary states, $\vec{\Psi}(t)\equiv\vec{\Psi}(0)e^{-i\Omega t}$, and $N_{R1}$, we have to search for the real spectrum, i.e.,
\begin{align}
  \Omega_{\pm}
  &=
  \frac{1}{2}
  \left(
  E_{1}
  +
  E_{2}
  \right)
  \pm
  \frac{1}{2}
  \sqrt{
  \left(
  E_{1}
  -
  E_{2}
  \right)^2
  +
  4J^2
  }~~
  \in
  \mathbb{R}
  \,.\label{Stationary eigen-energy}
\end{align}
The stationary states pumped from one side must possess a finite d.c. Josephson current $2J\sqrt{N_{c1}N_{c2}}\sin(\Delta\varphi)$ to balance the loss from the other side, and the relative phase $\Delta\varphi\equiv\varphi_{2}-\varphi_{1}$ across the BJJ (with respect to $\Omega_{-}$ and $\Omega_{+}$) deviates from 0 and $\pi$, corresponding to the usual bonding and anti-bonding states with zero current.

\begin{figure*}
  \includegraphics[width=0.325\linewidth]{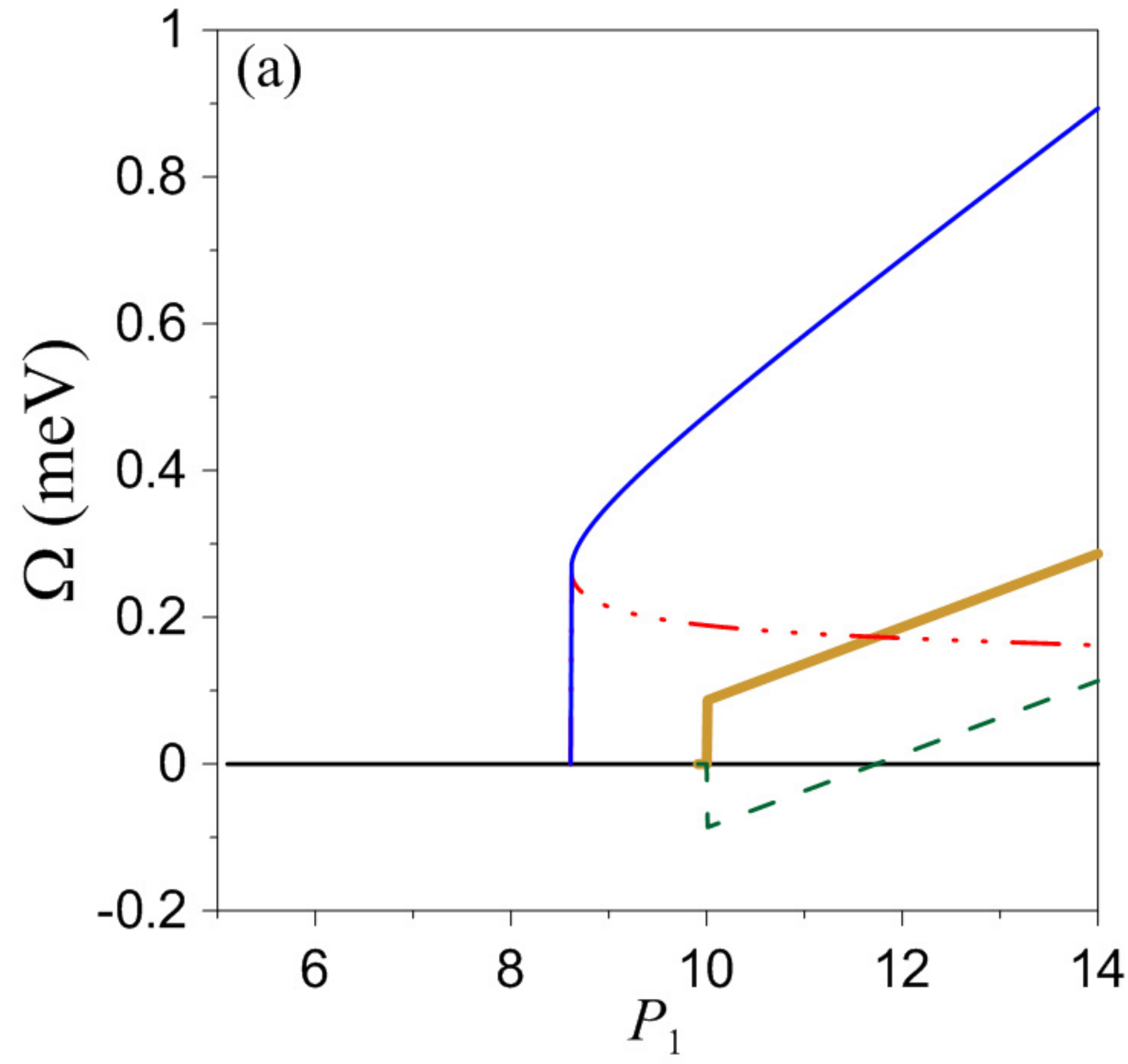}
  \includegraphics[width=0.325\linewidth]{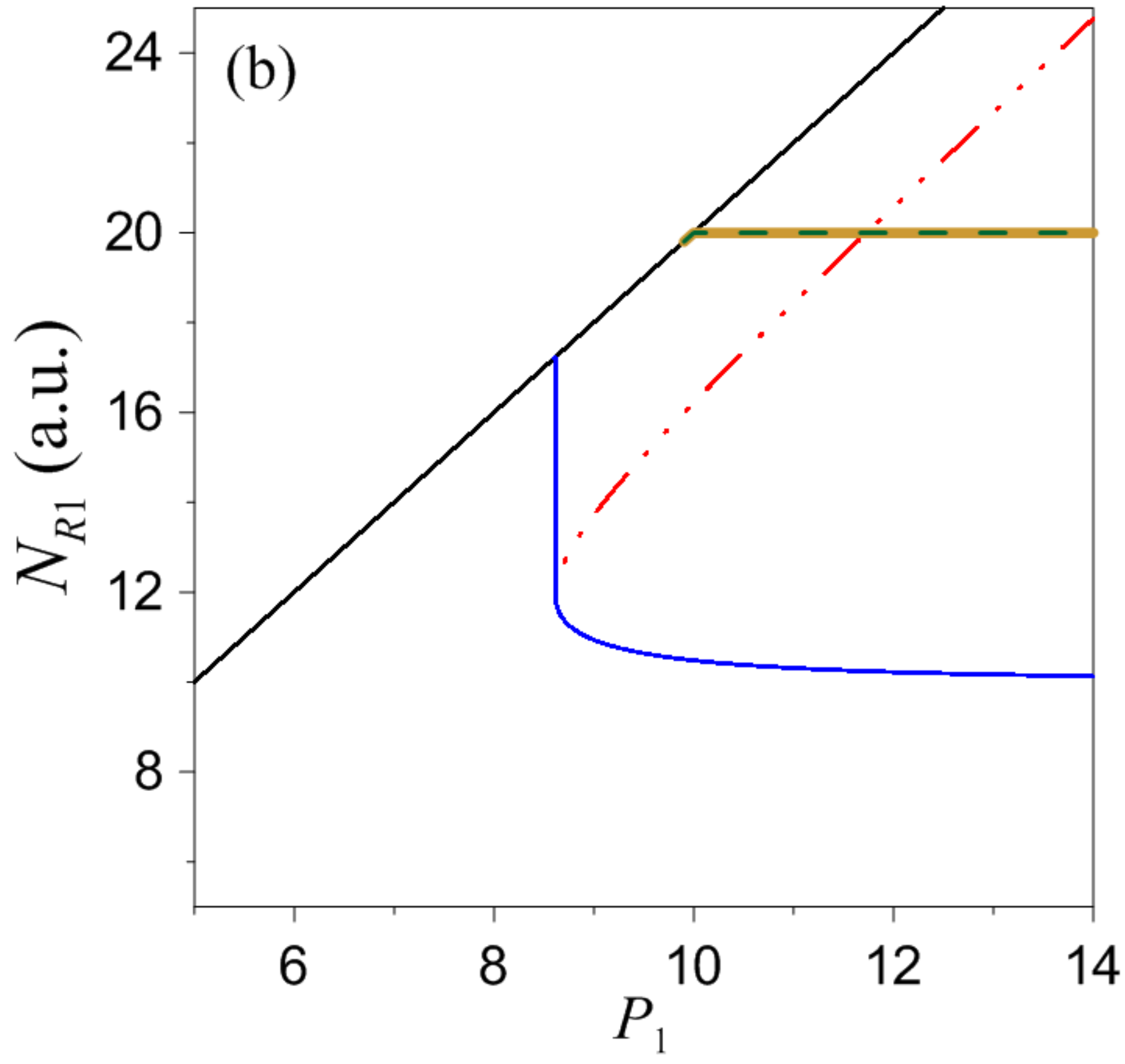}
  \includegraphics[width=0.325\linewidth]{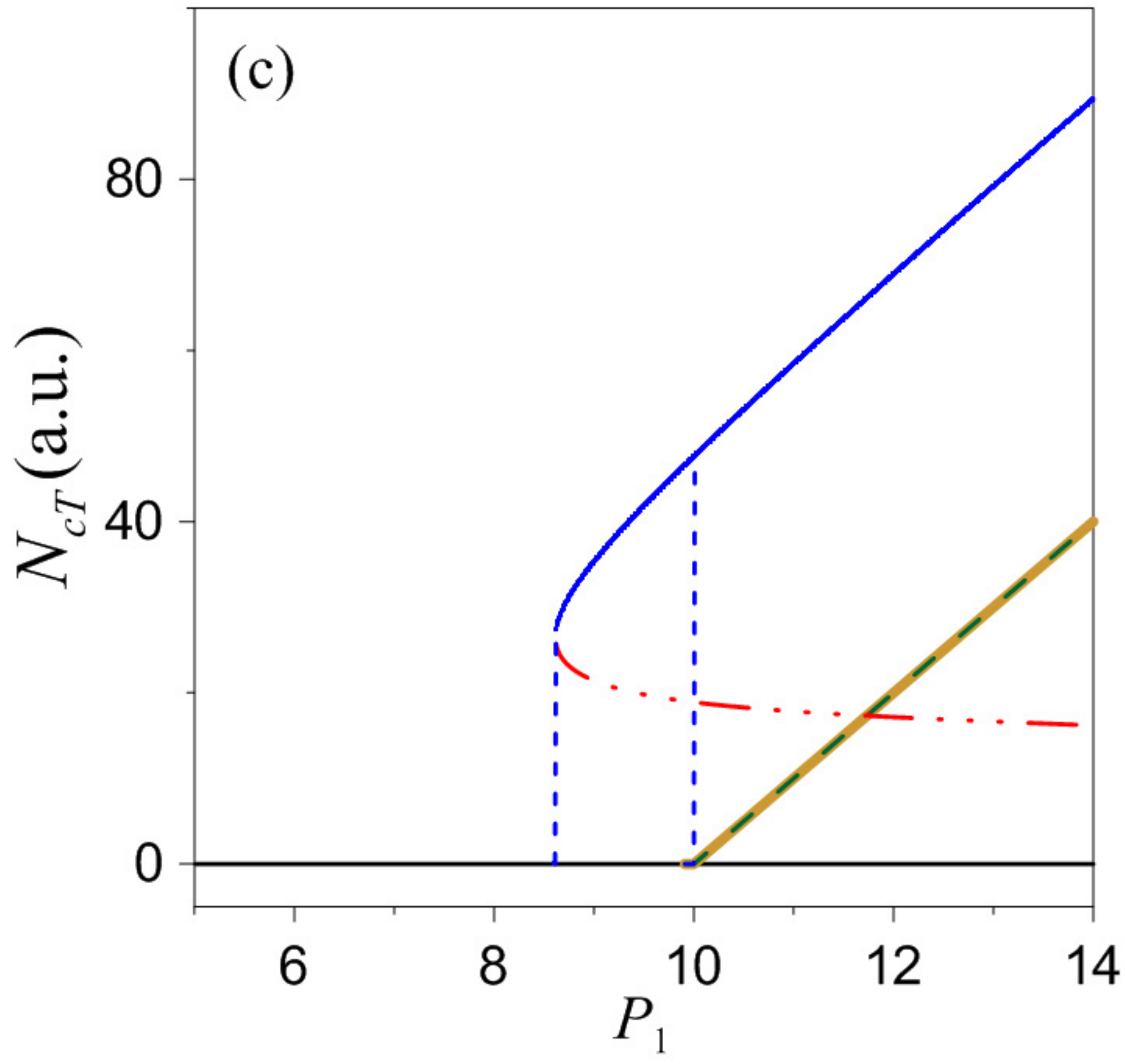}
  \includegraphics[width=0.325\linewidth]{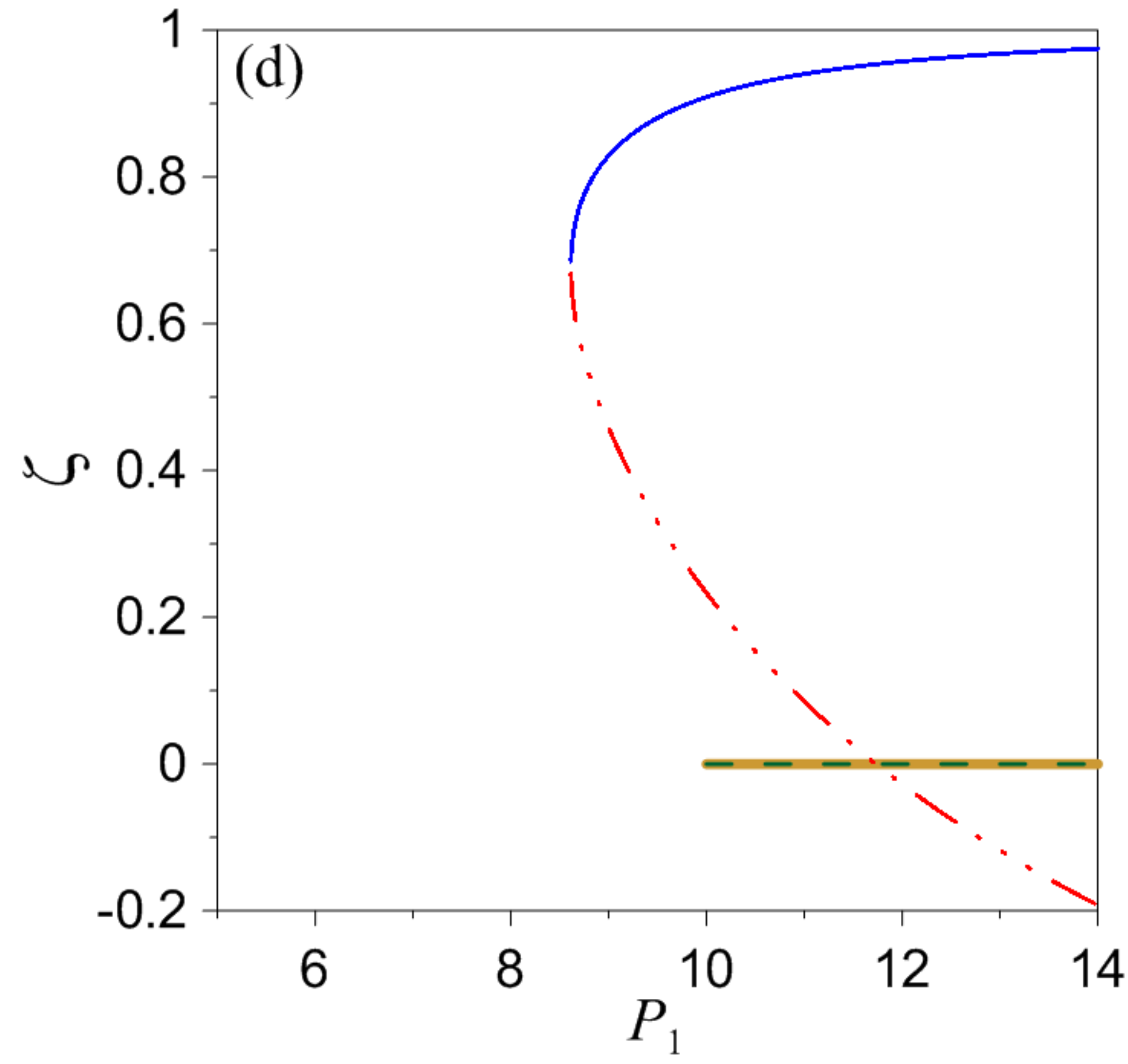}
  \includegraphics[width=0.33\linewidth]{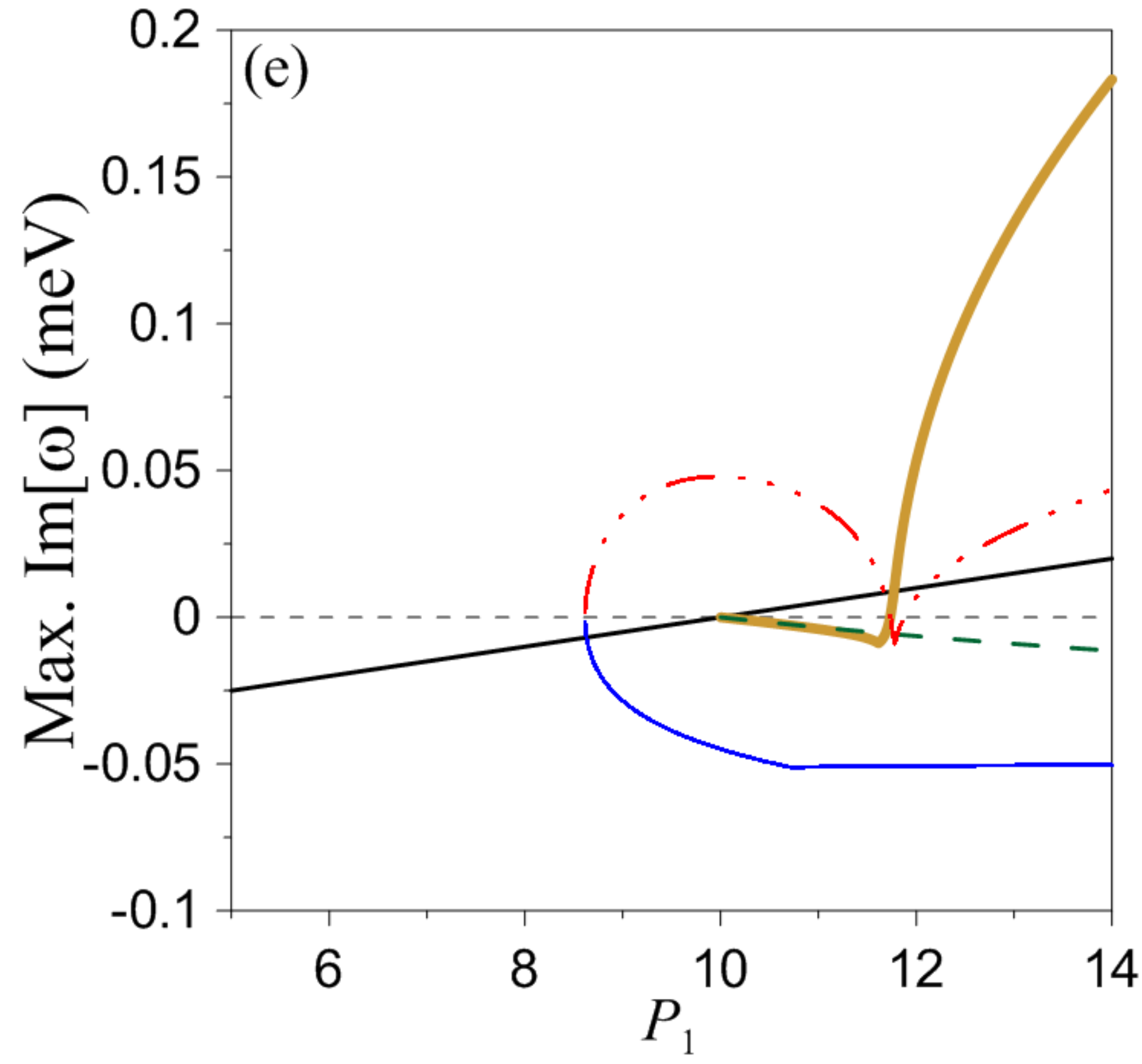}
  \includegraphics[width=0.325\linewidth]{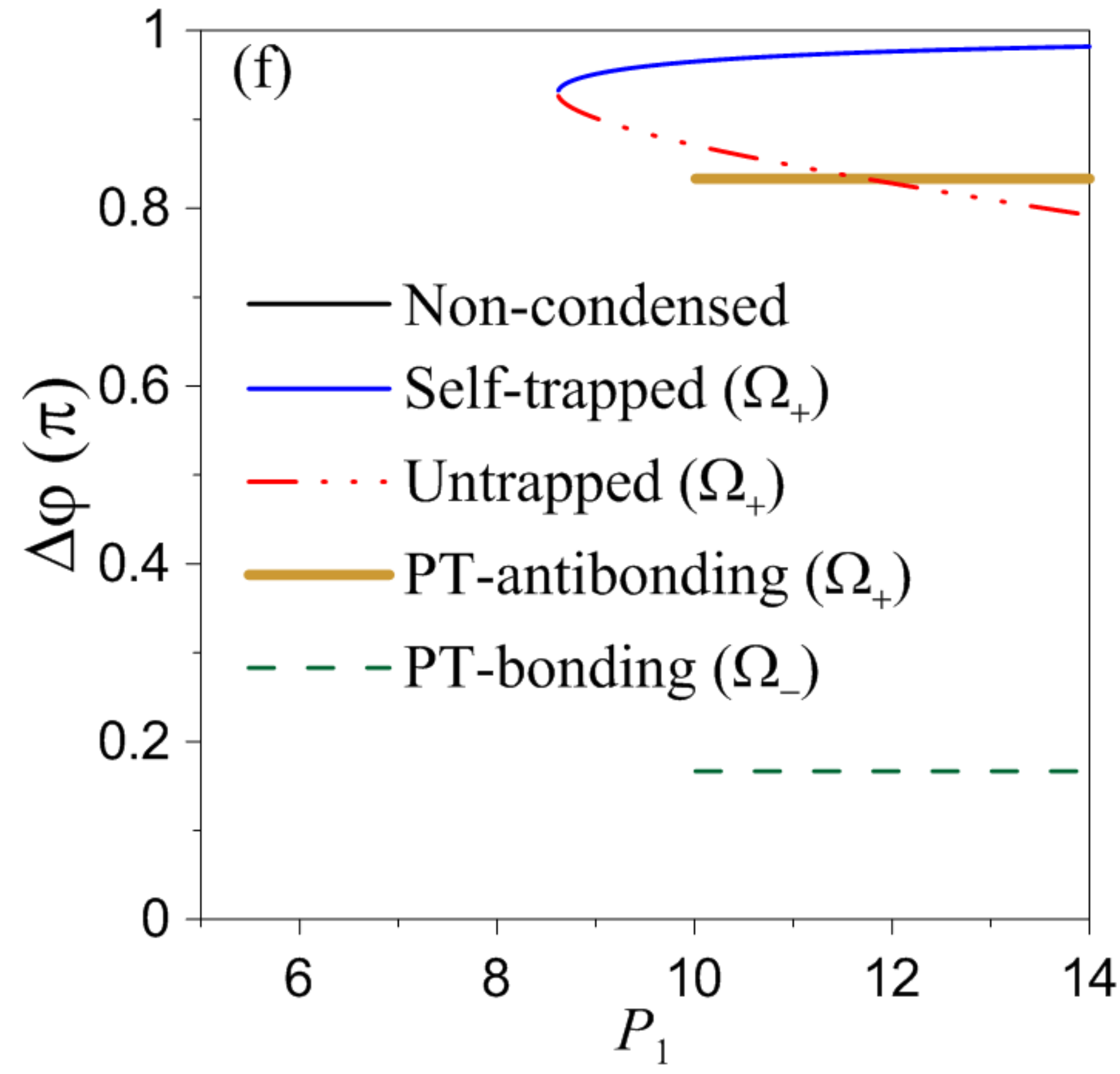}
  \caption{(Color online)
  Signatures of four stationary condensate states with respect to the pumping strength $P_{1}$: 
  (a) Energy spectrum corresponding to the real eigenvalues $\Omega_{+}$ or $\Omega_{-}$; 
  (b) reservoir-polariton number $N_{R1}$ in response to the condensates; 
  (c) total condensate-polariton number $N_{cT}$; 
  (d) population imbalance $\zeta$; 
  (e) stability determined by the maximum imaginary part of the fluctuation spectrum; 
  and 
  (f) phase difference $\Delta\varphi$. 
  These states are not necessarily orthogonal with each other due to the nonlinearity, and they can be classified into two groups.
  The solid blue and dash-dotted red curves represent the asymmetric (self-trapped and untrapped) states, and both of them are anti-bonding states ($\Omega_{+}$) with phase difference close to $\pi$.
  The thick yellow and dashed green curves correspond to the PT-symmetric states with $\Omega_{+}$ and $\Omega_{-}$, respectively, and they coincide in (b-d) except having different energy and phase.
  Above the threshold, two (for large pumping) to three (for small pumping) states are stable.
  Below the threshold there exists a hysteresis behavior between the self-trapped state and the non-condensed state, shown by the blue-dashed vertical lines in (c).
  The threshold pumping $P_{\textnormal{th}}=10$ is determined by the non-condensed state, which is shown by the black lines [not shown in (d) and (f)] and becomes unstable above the threshold.
  The parameters used are $\epsilon_{1}=\epsilon_{2}=0$, $U_{1}=U_{2}=0.01$ meV, $\gamma_{1}=\gamma_{2}=0.1$ meV, $\gamma_{R1}=0.5$ meV, $R_{1}'=0.01$ meV, and $J=0.1$ meV. 
  }\label{Stationary states & multi-stability}
\end{figure*}

Two analytic solutions can be derived from $R_{1}(N_{R1})=\gamma_{1}+\gamma_{2}$; one corresponds to $\Omega_{+}$ and the other to $\Omega_{-}$ (see yellow and green curves in Fig.~\ref{Stationary states & multi-stability}).
In this case, the Hamiltonian possesses parity-time (PT) symmetry, i.e. $E_{1}=E_{2}^*$, where the injection of the condensate polaritons at site-1, $R_{1}(N_{R1})-\gamma_{1}$, is equal to the decay at site-2.
\cite{Graefe:J.Phys.A:45.444015(2012)(PT-symmetry_BEC),Cartarius:PRA.86.013612(2012)(PT-symmetric_BEC_delta-function_double_well)} 
These stationary states only exist with a real spectrum under the condition 
\begin{align}
  J^2
  \geq
  \frac{1}{16}
  \left[
  R_{1}(N_{R1})
  -
  \gamma_{1}
  +
  \gamma_{2}
  \right]^2
  =
  \frac{\gamma_{2}^2}{4}
  \,.\label{Spontaneous PT-symmetry breaking}
\end{align}
Violating this condition by increasing the decay or decreasing the tunneling makes the spectrum complex and leads to \textit{spontaneous PT-symmetry breaking}.
\cite{Cartarius:PRA.86.013612(2012)(PT-symmetric_BEC_delta-function_double_well),C.M.Bender:Rep.Prog.Phys.70.947(2007)(review_non-Hermitian_Hamiltonians),Peng&Nori&Bender:NPhy.10.394(2014)(PT-symmetric_whispering-gallery_microcavities),Peng&Ozdemir&Bender&Nori:Science.346.328(2014)(Loss-induced_suppression_revival_lasing),Jing&Ozdemir&Lu&Zhang&Yang&Nori:PRL.113.053604(2014)(PT-symmetric_phonon_laser)} 
The equal sign of Eq.~(\ref{Spontaneous PT-symmetry breaking}) gives the exceptional point, where these two states coalesce.

For both PT-symmetric states, the condensate polaritons are equally populated across the junction,
\begin{align}
  |\Psi_{1}|^2
  =
  |\Psi_{2}|^2
  &=
  \frac{P_{1}-\gamma_{R1}N_{R1}}{R_{1}(N_{R1})}
  \,.
\end{align}
By coincidence,
these states are created above the threshold because the reservoir polaritons have to be pumped to the condition $N_{R1}=N_{R\textnormal{th}}$. 
The reservoir-polariton number is kept constant above the threshold and the total condensate-polariton number ($N_{cT}\equiv N_{c1}+N_{c2}$) increases linearly with the pumping [Fig.~\ref{Stationary states & multi-stability}(b, c)].

In general, there exist two other solutions, both corresponding to $\Omega_{+}$, with the imbalanced population of the condensate polaritons $\zeta\equiv(N_{c1}-N_{c2})/N_{cT}\neq0$ [Fig.~\ref{Stationary states & multi-stability}(d)]. 
They are obtained by numerically finding the roots of $\textnormal{Im}(\Omega)=0$.
One solution is localized in site-1 (the blue curves in Fig.~\ref{Stationary states & multi-stability}), and this localization is due to the same mechanism of macroscopic quantum self-trapping for short-time oscillations,
\cite{Raghavan:PRA.59.620(1999)(BEC_Josephson_effects)} 
where the interaction between the condensate polaritons shifts the energy difference across the BJJ and reduces the Josephson current.
Therefore, increasing the pumping or decreasing the junction tunneling enhances the self-trapping effect.
The other solution is more populated in site-2 with weaker condensation (the red curves in Fig.~\ref{Stationary states & multi-stability}) and the self-trapping effect is limited.
Increasing the pumping will drive more polaritons tunneling to the unpumped site and further reduce the condensation.
Thus, the solution reduces to the zero-condensate state above a critical pumping. 
Interestingly, the imbalanced states appear even below the threshold. 
We will show later that the polariton BJJ exhibits bistability below the threshold.


\subsection{Multiple stability}

In order to determine which states can be observed experimentally, the stability is analyzed by calculating the complex spectrum of deviation from stationary states.  However, we do not employ the usual derivation of the elementary excitations by directly linearizing Eq.~(\ref{GGPE}), which includes both the phase terms $\varphi_{1}$ and $\varphi_{2}$. 
This is because only the phase difference 
is physically meaningful without considering the local dispersion. Instead, we derive the equations of motion of the phase difference $\Delta\varphi$ and the population imbalance $\zeta$, 
similar to the conserved BEC systems,
\cite{Raghavan:PRA.59.620(1999)(BEC_Josephson_effects)} 
and take into account the additional degrees of freedom, i.e. the total condensate-polariton number $N_{cT}$ and the reservoir-polariton number $N_{R1}$. 
The equations of motions are given by
\begin{align}
  \dot{\zeta}
  &=
  V_{12}^{I}
  \left(
  1-\zeta^2
  \right)
  -
  2J
  \sqrt{1-\zeta^2}
  \sin(\Delta\varphi)
  \nonumber\\
  \Delta\dot{\varphi}
  &=
  \epsilon_{12}
  +
  V^{R}_{12}
  +
  \left(
  {
  U_{12}
  \over
  2
  }
  +
  \bar{U}
  \zeta
  \right)
  N_{cT}
  \nonumber
  \\&
  +
  2J
  {
  \zeta
  \over
  \sqrt{1-\zeta^2}
  }
  \cos\left(\Delta\varphi\right)
  \label{Non-equilibrium eq of motion of polariton BJJ}
  \\
  \dot{N}_{cT}
  &=
  \left[
  2
  \bar{V}^{I}
  +
  V_{12}^{I}
  \zeta
  \right]
  N_{cT}
  \nonumber\\
  \dot{N}_{R1}
  &=
  P_{1}
  -
  \gamma_{R1}
  N_{R1}
  -
  R_{1}(N_{R1})
  N_{cT}
  {1+\zeta\over2}
  \,,
  \nonumber
\end{align}
where
\begin{align}
  \bar{U}
  &\equiv
  \frac{U_{1}+U_{2}}{2}
  \nonumber\\
  U_{12}
  &\equiv
  U_{1}-U_{2}
\end{align}
and similar definitions are applied to $V_{j}^{R/I}$ and $\epsilon_{j}$.
The term $\epsilon_{12}+V_{12}^{R}$ corresponds to the effective detuning including the reservoir polaritons and high-energy excitons.
By linearizing Eq.~(\ref{Non-equilibrium eq of motion of polariton BJJ}),
we can calculate the fluctuation spectrum and analyze the dynamical stability with respect to the stationary states.

Figure \ref{Stationary states & multi-stability}(e) shows the maximum imaginary part of the fluctuation spectrum. 
One of the PT-symmetric states $\Omega_{-}$ is always stable and the other $\Omega_{+}$ becomes unstable when increasing $P_{1}$. 
As for the imbalanced states, the self-trapped state is stable except for a small region below the threshold, and the other one (untrapped state) is generally unstable unless it is close to the PT state with $\Omega_{+}$. 
This leads to multi-stability of the condensation resulting from the polariton tunneling. 

Below threshold, there exists a bistable regime with a self-trapped condensed state and a non-condensed state.
The bistability can be explained by damped oscillations of the polariton BJJ.
The self-trapping oscillations occur if the initial polariton number is larger than a critical value [$\bar{U}N_{cT}(0)/2J>\Lambda_{c}$], with a suitable range of the initial imbalance $\zeta(0)$,
\cite{Raghavan:PRA.59.620(1999)(BEC_Josephson_effects)} 
and it is eventually damped to the imbalanced equilibrium position $\zeta(t\rightarrow\infty)>0$.
Otherwise, the oscillations are damped to a balanced state with $\zeta(t\rightarrow\infty)=0$, and from the condition $\dot{N}_{cT}=0$ in 
(\ref{Non-equilibrium eq of motion of polariton BJJ}) 
the condensation must vanish unless the threshold is reached.
If the pumping strength is increased from zero, the non-condensed state holds without initial condensation and becomes unstable across the threshold.
The self-trapped state dominates just above the threshold due to the minute occupation of the other states.
After then, the above-critical polariton number keeps the condensation when the pumping decreases.
Thus the hysteresis of condensation could be experimentally observed by cyclically increasing and decreasing the pumping near the threshold, provided that the fluctuation of pumping intensity is reduced, e.g., by employing a cw diode laser.
\cite{Love:PRL.101.067404(2008)(Intrinsic_decoherence_coexisting_states_polariton_BEC),Krizhanovskii:PRB.80.045317(2009)(Coexisting_states_disordered_polariton_BECs)}

Above the threshold, the multiple stable states are also determined by the initial values. 
When the self-trapping condition is not satisfied, the condensates evolve to other stable states.
In conserved BJJs,
\cite{Raghavan:PRA.59.620(1999)(BEC_Josephson_effects)} 
for $|\Delta\varphi(0)|\leq\pi/2$, the self trapping occurs in the running-phase modes provided $\zeta(0)>\zeta_{c}$, while for $|\Delta\varphi(0)|>\pi/2$, it could be either in the running-phase modes or the $\pi$-phase modes, depending on both $N_{cT}(0)$ and $\zeta(0)$.
The suitable range of $\zeta(0)$ is altered by the reservoir and the decay, and a large initial imbalance generally induces the self trapping.
We have shown that, for zero detuning, the self-trapped state must be antibonding-like with a phase difference close to $\pi$, even if the initial phase is zero [Fig.~\ref{Stationary states & multi-stability}(f)].
This can be understood by the running phase being damped to $(2n+1)\pi$ and eventually reaching the antibonding-like stationary state.
One notes that the damped running phase and $\pi$-phase locking have been observed in a recent experiment.
\cite{Abbarchi:NPhy(2013)(Polariton_BEC_Josephson_oscillations)} 
This phenomenon seems to be different from our case because the eigen-energy is never real without pumping, i.e., no stationary condensate is achieved.
We also derive an oscillator model to explain the $\pi$-phase locking either with or without the stationary condensation 
in the next section.

In order to experimentally observe the behavior of each stable state, an additional pulsed laser resonant with the polariton energy band is necessary to coherently excite the system, before the incoherent pumping is applied.
The initial conditions of the condensate polaritons can be well controlled by exploiting this technique, where an arbitrary population imbalance in a linear combination of bonding and antibonding states can be created.
\cite{Abbarchi:NPhy(2013)(Polariton_BEC_Josephson_oscillations)}
For comparison, another system, in which the incoherent pumping is replaced by a cw resonant laser, is discussed here.
Under the long-time coherent pumping, the dynamics is essentially different from our study. 
No reservoir polaritons are excited and the stationary states depend on the laser frequency.
\cite{Carusotto&Ciuti:RevModPhys.85.299(2013)(Quantum_fluids_light)}
In this case, a regime of optical bistability exists for a single condensate with the pumping frequency larger than the ground-state energy of polaritons.
\cite{Baas:PRA.69.023809(2004)(Optical_bistability-coherent_pumping_polariton_BEC)}
The tunneling of a polariton BJJ, under one-side pumping, further gives rise to several unstable regimes for large pumping amplitudes.
\cite{Sarchi&Carusotto&Wouters&Savona:PRB.77.125324(2008)(Coherent_driving_instabilities_polariton_BJJ)}
However, in our study, there exists at least one stable state (zero-condensate or self-trapping  state) for all pumping strength.

\section{Oscillator model}


\subsection{Self-trapping regime}\label{Pi-phase_locking analysis}

\begin{figure}[]
  \includegraphics[width=\linewidth]{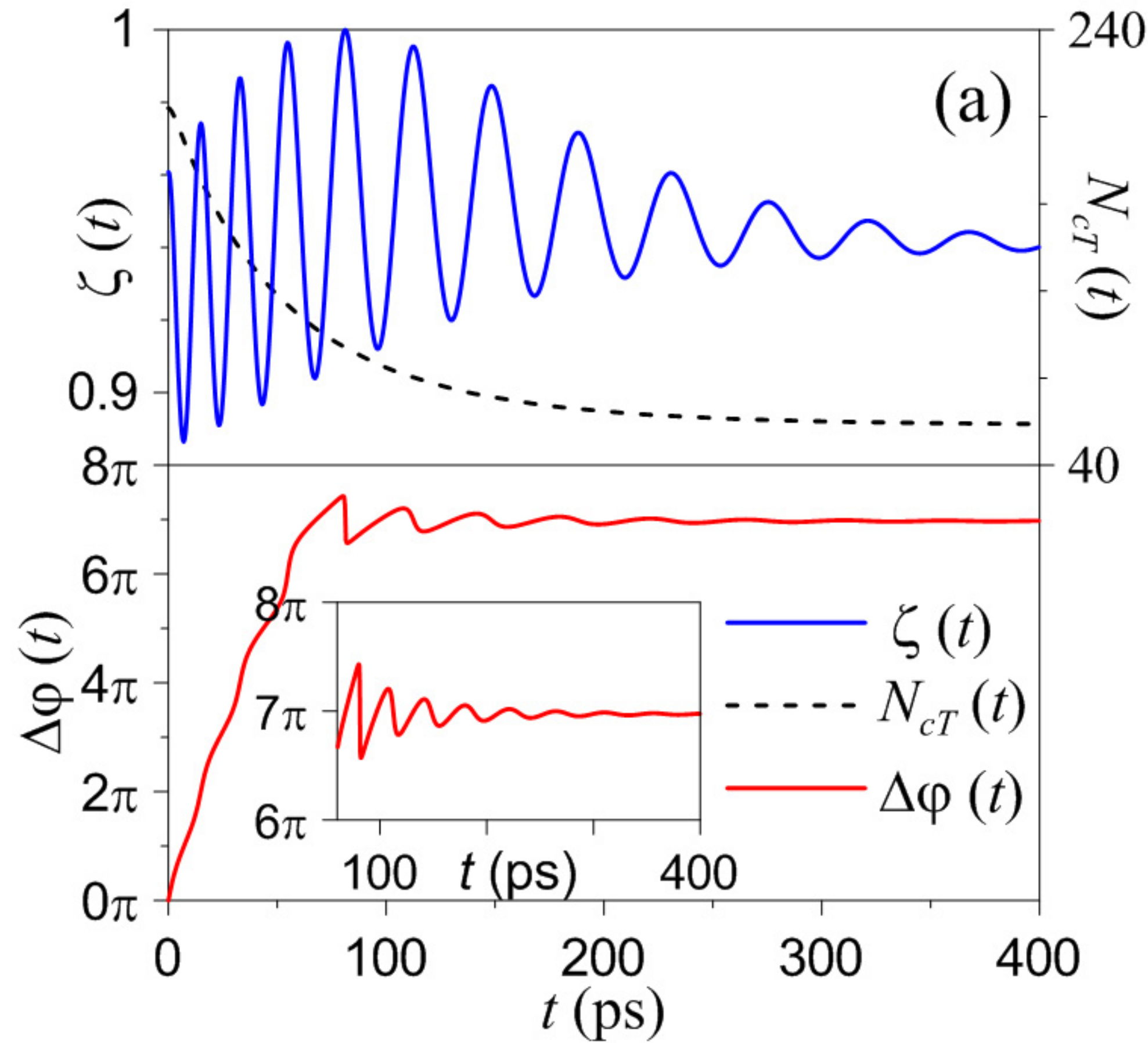}
  \includegraphics[width=\linewidth]{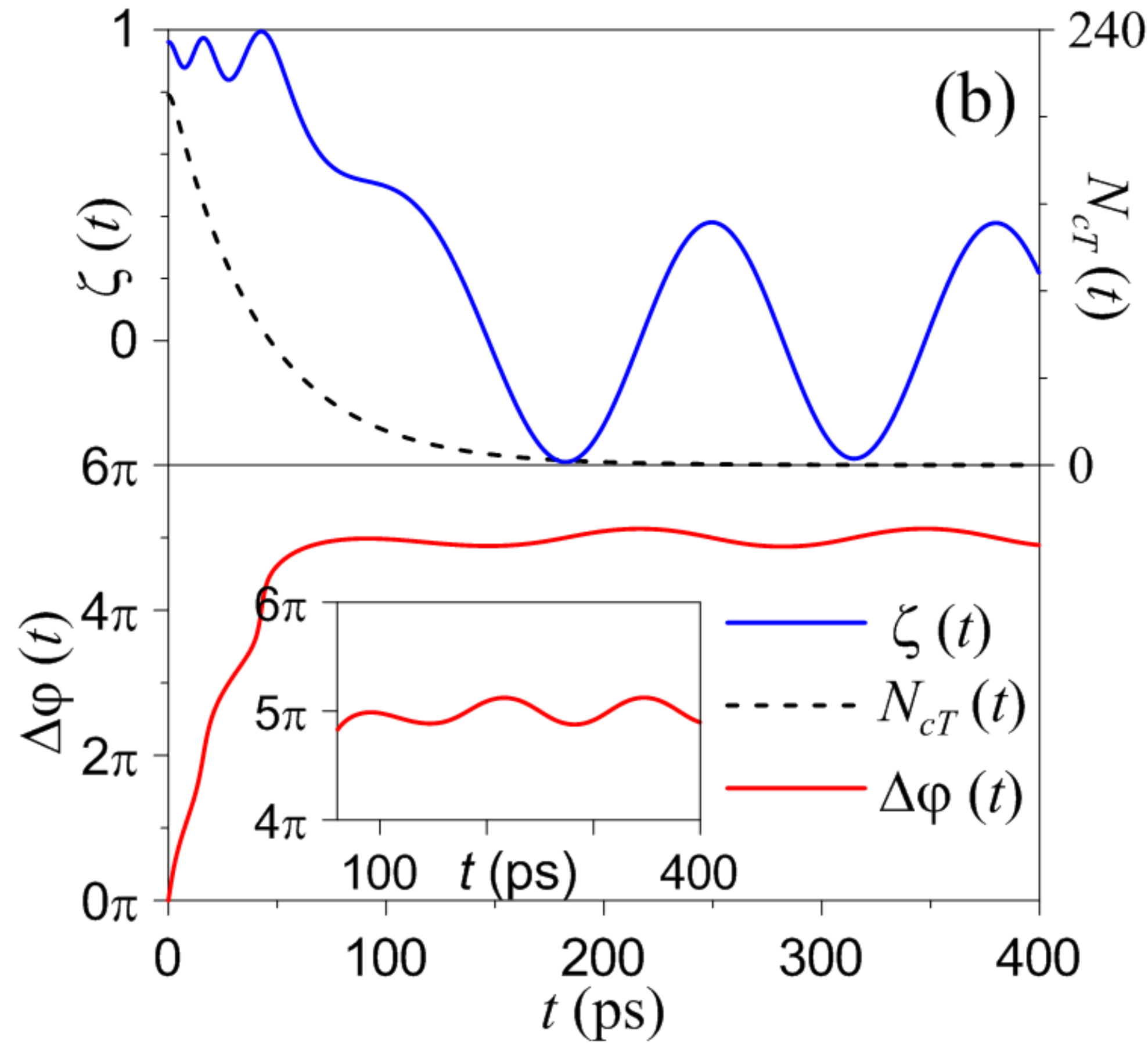}
  \caption{(Color online)
  Time dependence of the imbalance $\zeta(t)$, the condensate-polariton number $N_{cT}(t)$, and the phase difference $\Delta\varphi(t)$ of the self-trapped state with zero initial phase difference for $P_{1}=11$ (a) and $P_{1}=0$ (b).
  The insets show the $\pi$-phase locking for $\Delta\varphi(t\rightarrow\infty)$.
  }\label{Pi-phase_locking}
\end{figure}

Figure \ref{Pi-phase_locking} shows, for zero detuning, the $\pi$-phase locking of the self-trapped states in a non-equilibrium polariton BJJ by solving Eq.~(\ref{Non-equilibrium eq of motion of polariton BJJ}), either with or without stationary condensation.
For the non-condensed case, our results agree well with the experiment.
\cite{Abbarchi:NPhy(2013)(Polariton_BEC_Josephson_oscillations)} 
It has been predicted that the imbalance of self-trapping eventually evolves into an oscillatory regime after the polariton number drops below the critical value.
\cite{Shelykh:PRB.78.041302(2008)(Polariton_BJJ_spinor_separation)}
However, the mechanism of $\pi$-phase locking is not understood yet.

Here, we derive a nonlinear dissipated-oscillator model from (\ref{Non-equilibrium eq of motion of polariton BJJ}) with, $U_{12}=0$, in order to further understand the mechanism of this phenomenon.
Under the self-trapping conditions, $\eta\equiv\sqrt{1-\zeta^2(t)}\ll 1$ and $\zeta(t)\approx1-\eta>0$, the system can reduce to a second-order differential equation
\begin{align}
  \Delta\ddot{\varphi}(t)
  &\approx
  \bar{U}
  \dot{N}_{cT}(t)
  -
  2J
  \left(
  {1-\eta\over\eta}
  \right)
  \sin\left[\Delta\varphi(t)\right]
  \Delta\dot{\varphi}(t)
  \nonumber\\
  &
  -
  2J^2
  \left(
  {1\over\eta}
  \right)
  \sin[2\Delta\varphi(t)]
  +
  O(\eta)
\end{align}
corresponding to a pendulum with a position-dependent dissipation and a decaying driving force $\bar{U}\dot{N}_{cT}$ satisfying
\begin{align}
  \bar{U}
  \dot{N}_{cT}
  &\approx
  \left[
  2
  \bar{V}^{I}
  +
  V_{12}^{I}
  (1-\eta)
  \right]
  \bar{U}
  N_{cT}
  \,.
\end{align}

The angle of the pendulum is defined by $2\Delta\varphi(t)$, and thus the pendulum has minimal potential energy for $\Delta\varphi=n\pi$, with $n$ being either odd or even.
The angular velocity is given by
\begin{align}
  \Delta\dot{\varphi}
  &=
  \Delta E
  +
  2J
  \left(
  {
  1-\eta
  \over
  \eta
  }
  \right)
  \cos\left(\Delta\varphi\right)
  \,,
  \label{Angular velocity for self-trapping}
\end{align}
where
\begin{align}
  \Delta E
  &\equiv
  \epsilon_{12}
  +
  V^{R}_{12}
  +
  \bar{U}
  (1-\eta)
  N_{cT}
  \,.
\end{align}
The angular velocity can be zero only for $\Delta\varphi\in(\pi/2,3\pi/2)$, 
because $\Delta E>0$ for zero detuning ($\epsilon_{12}+V_{12}^{R}=0$).
Hence the pendulum is stable for odd $n$.
If the initial angular velocity is small, the pendulum oscillates with small amplitude around the $\pi$ phase.
Otherwise, it moves with an increasing phase. 
The driving force is able to decelerate the running phase and to lock the $\pi$ phase before $\bar{U}\dot{N}_{cT}\approx0$, as long as $N_{cT}(0)$ is large enough.

This model can be also extended to the regime of nonzero detuning.
From the condition $\Delta E>0$ in Eq.~(\ref{Angular velocity for self-trapping}), the $\pi$-phase locking holds for positive detuning and small negative detuning, i.e., $\epsilon_{12}+V_{12}^{R}>-\bar{U}N_{cT}$.
Otherwise, the phase will be locked around zero for large negative detuning.
Thus, we can unify the phase-locking phenomenon for both the condensed and non-condensed cases.
A main difference between these two cases is the population imbalance after the phase is locked.
For the non-condensed case, $N_{cT}$ decays to zero and the self-trapping ($\eta\ll 1$) no longer holds.
However, the phase is still locked because the angular velocity has been decelerated to be small.

\subsection{Josephson regime}\label{Josephson regime}

In contrast to the self-trapping regime, the dynamics of a.c.~Josephson oscillations can be investigated in the limit of $|\zeta|\ll1$. 
In the Josephson regime, i.e., $\bar{U}N_{cT}\gg J$, the equations of motion (\ref{Non-equilibrium eq of motion of polariton BJJ}) can be simplified as
\begin{gather}
  \dot{\zeta}
  \approx
  V_{12}^{I}
  -
  2J
  \sin(\Delta\varphi)
  \,,
  \nonumber\\
  \Delta\dot{\varphi}
  \approx
  \epsilon_{12}
  +
  V_{12}^{R}
  +
  \bar{U}
  N_{cT}
  \zeta
  \,,
  \label{Josephson eqs. with current source}
\end{gather}
with $\dot{N}_{cT}\approx0$ and $\dot{N}_{R1}\approx0$.
This model mimics a superconducting Josephson junction applied by an extra d.c.~current source $V_{12}^{I}>0$ due to the one-side pumping.
For zero detuning $\epsilon_{12}+V_{12}^{R}=0$, the population is balanced and the states possess the PT symmetry, as long as the current source is smaller than the critical current $2J$.
This corresponds to the d.c.~Josephson effect, where the phase differences for bonding and antibonding states are determined by the ratio of $V_{12}^{I}$ and $2J$, with $0<\sin(\Delta\varphi)\leq1$.
If $V_{12}^{I}>2J$, the population imbalance will deviate from the limit of $|\zeta|\ll1$ by charging across the junction, and eventually evolve to the self-trapped state.
This is exactly the spontaneous PT-symmetry breaking from Eq.~(\ref{Spontaneous PT-symmetry breaking}).

For the cases of equivalent pumping ($V_{12}^{I}=0$),
\cite{Wouters:PRB.77.121302(2008)(Synchronization_phase_transition_polariton_BJJ),Borgh:PRB.81.235302(2010)(Spinor_polariton_BEC)} 
the condensates are desynchronized when the detuning is larger than a critical value, resembling the a.c.~Josephson effect driven by a bias voltage.
However, the d.c.~current source results in a finite average charging rate $\langle\dot{\zeta}(t)\rangle>0$, which either balances the bias voltage to lock the running phase in the Josephson regime, or drives the population into the self-trapping regime.
Therefore, the a.c.~oscillations in our study are washed out by the d.c.~current drive, and the condensates are always synchronized.
Although the stationary population is asymmetric in the presence of detuning, 
a phenomenon similar to the spontaneous PT-symmetry breaking can still take place since the charging rate $\dot{\zeta}$ is dominated by the competition between $V_{12}^{I}$ and $2J$.


\section{Discussion}

\subsection{Detuning effects}\label{Detuning}

\begin{figure*}
  \includegraphics[width=0.49\linewidth]{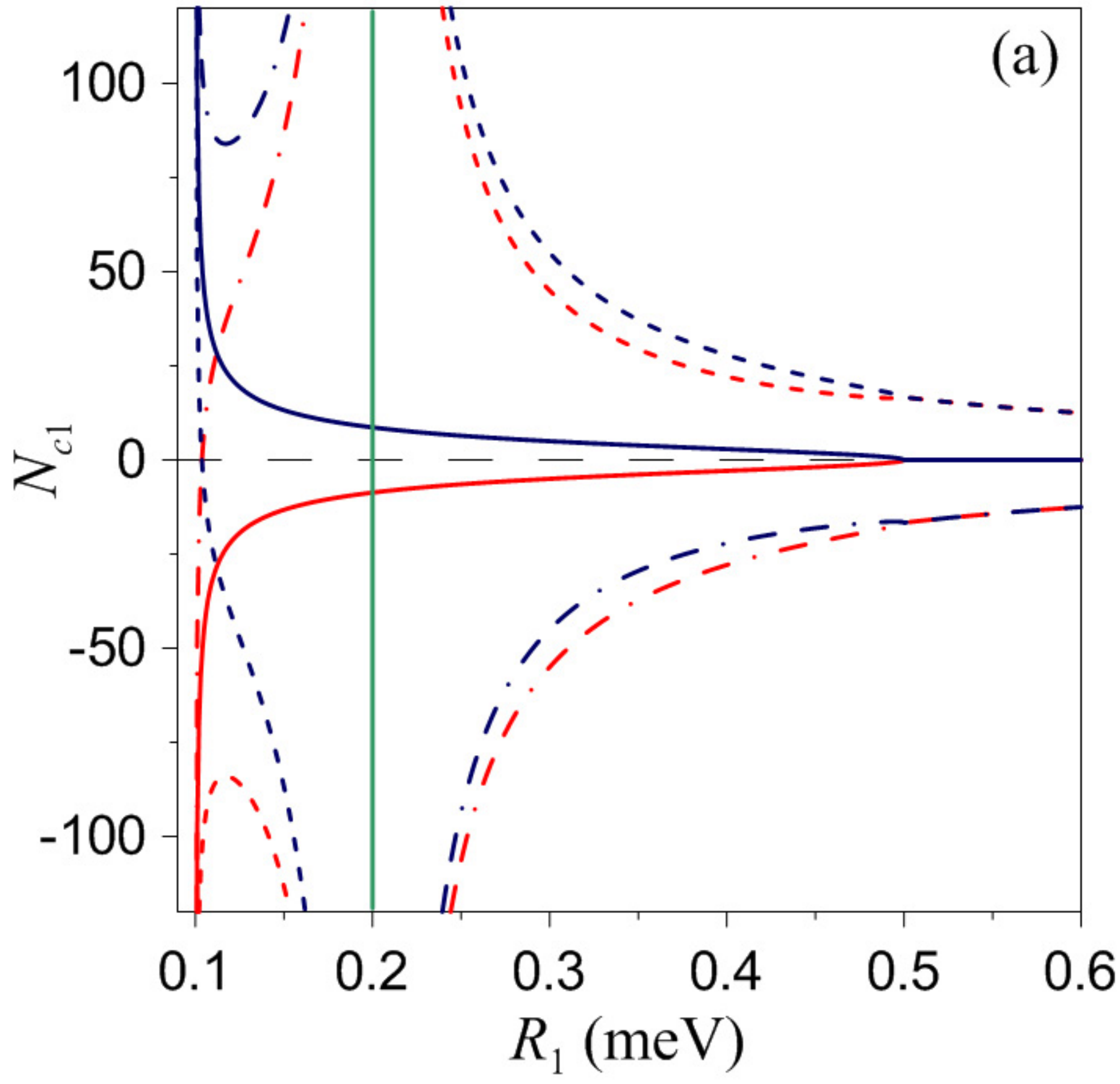}
  \includegraphics[width=0.49\linewidth]{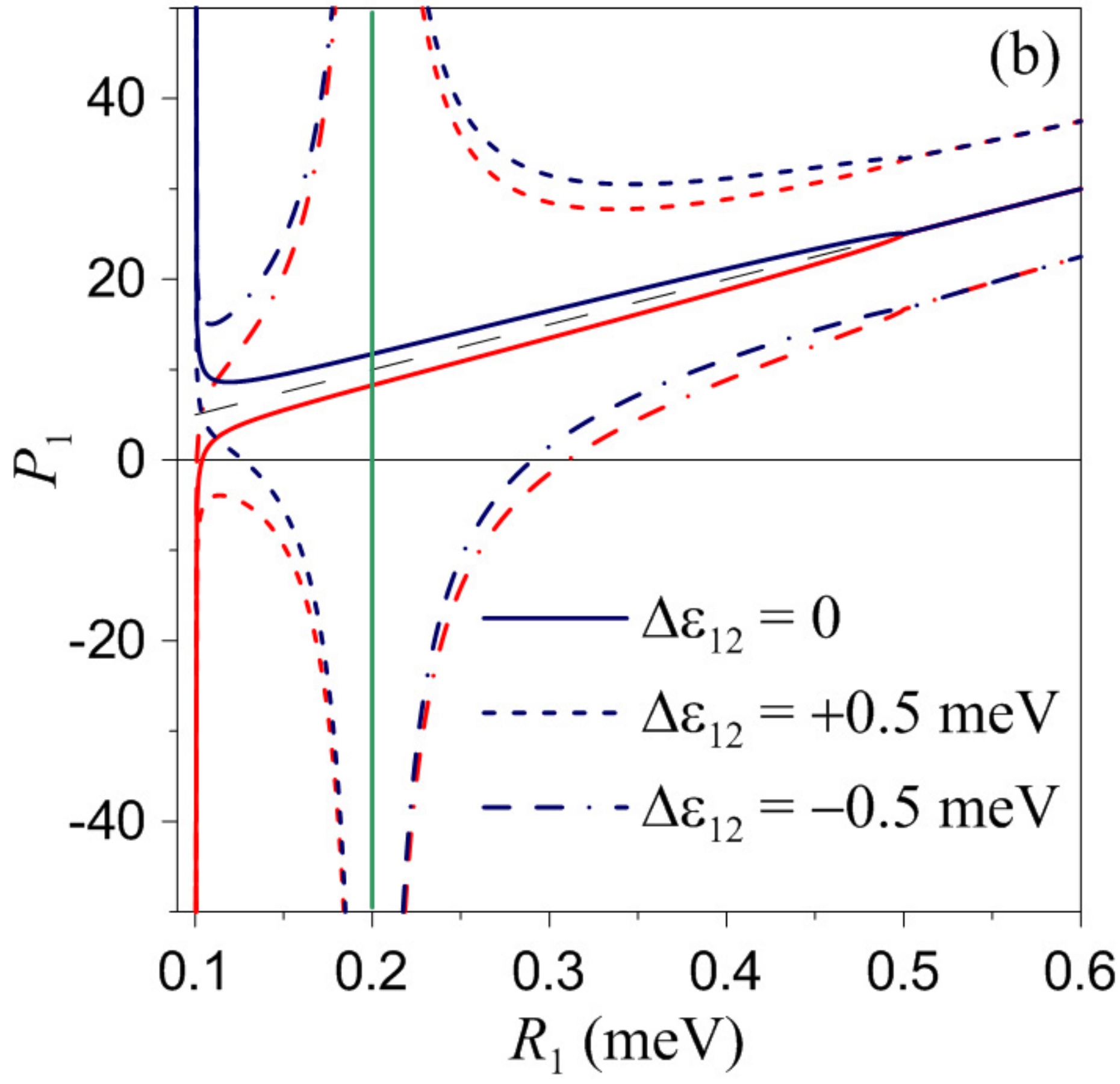}
  \caption{(Color online)
  Detuning effects of the stationary states: 
  (a) The number $N_{c1}$ of condensate polaritons at the pumped side for the stationary solutions with $R_{1}\geq\gamma_{1}=0.1$ meV and (b) the corresponding pumping $P_{1}$, both as functions of the stimulated scattering $R_{1}$ for three different detunings $\Delta\varepsilon_{12}\equiv\epsilon_{12}+V_{12}^{R}$.
  The dark blue and red curves represent antibonding ($\Omega_{+}$) and bonding ($\Omega_{-}$) asymmetric states, respectively. 
  The green vertical line corresponds to both the PT-symmetric states ($\Omega_{\pm}$) for zero detuning.
  These solutions make sense only for $N_{c1}\geq0$ and $P_{1}\geq0$, i.e., the upper half-plane in (a) and the area above the thin dashed line (corresponding to $P_{1}=\gamma_{R1}N_{R1}$) in (b).
  The detuning introduces hybridization and anticrossing between the PT-symmetric and asymmetric states, and changes the dynamics for small pumping.
  However, increasing the pumping strength would restore the zero-detuning behavior, i.e., the bistability of the self-trapped state and the PT-symmetric bonding state.
  }\label{Detuning effects}
\end{figure*}

It is important to consider the interaction between the condensate and reservoir polaritons, as well as that between the condensate polaritons and high-energy excitons, which are responsible for, above threshold, the blue shift of the polariton energy band in the order of meVs.
\cite{Brichkin:PRB.84.195301(2011)(Interaction_between_reservoir&polariton_BEC)}
Under the one-side pumping, the blue shift of the pumped side induces an energy difference across the polariton BJJ.
On the other hand, a difference of the confinement potential between the pillars might take place during the fabrication process.
These effects contribute to the effective detuning term $\Delta\varepsilon_{12}\equiv\epsilon_{12}+V_{12}^{R}$ in Eq.~(\ref{Non-equilibrium eq of motion of polariton BJJ}).

In order to analyze the detuning effects, all particle numbers of the stationary states can be represented as functions of the stimulated scattering $R_{1}$, i.e.,
\begin{gather}
  N_{c1}
  =
  N_{c1}(R_{1})
  \geq0
  \nonumber
  \\
  N_{R1}(R_{1})
  =
  \frac{R_{1}}{R_{1}'}
  \geq0
  \label{particle numbers of R_1}
  \\
  N_{c2}(R_{1})
  =
  \frac{R_{1}-\gamma_{1}}{\gamma_{2}}
  N_{c1}(R_{1})
  \geq0
  \,,
  \nonumber
\end{gather}
and the corresponding pumping can be derived from Eq.~(\ref{reservoir rate eq.}), i.e.,
\begin{align}
  P_{1}(R_{1})
  &=
  \gamma_{R1}
  N_{R1}(R_{1})
  +
  R_{1}
  N_{c1}(R_{1})
  \geq0
  \,.
  \label{P_1 of R_1}
\end{align}
The condensate number of the first site $N_{c1}$ and the pumping strength $P_{1}$ are solved by $\Omega_{\pm}[N_{c1}(R_{1}),R_{1}]\in\mathbb{R}$ from Eq.~(\ref{Stationary eigen-energy}), as shown in Fig.~\ref{Detuning effects}.
In general, the states with $R_{1}\gg\gamma_{1}+\gamma_{2}$ (0.2 meV here) are unstable since they are close to the non-condensed states.

In the presence of detuning, the PT-symmetric states and the asymmetric states are hybridized near the threshold.
From Fig.~\ref{Threshold}(b), the threshold of a polariton BJJ reduces to that of a single condensate by increasing $|\Delta\varepsilon_{12}|$.
This is similar to the mechanism that the Rabi oscillations in a double-quantum-dot system are suppressed by detuning.
For positive detuning, the self-trapped state is still antibonding and appears at lower pumping strength (the dashed blue curve close to $R_{1}=0.1$ in Fig.~\ref{Detuning effects}).
The other states appear with a finite polariton number at larger pumping strength.
For negative detuning, the bonding state with zero phase dominates the self trapping near the threshold (the dash-dotted red curve close to $R_{1}=0.1$), and it becomes more PT-symmetric (close to $R_{1}=0.2$) when the the antibonding states (the dash-dotted blue curve) emerge with increasing $P_{1}$.
These are related to the previous discussed criterion $\Delta E>0$ of $\pi$-phase locking.
The interactions between condensate polaritons and incoherent polaritons (and excitons) contribute to a positive detuning ($V_{12}^{R}>0$).
Due to the reduced threshold, the bistability below $P_{\textrm{th}}$ could become insignificant.
However, a negative potential difference $\epsilon_{12}<0$ can recover this bistable regime.

At high condensate density, the hybridization becomes weak, and all the states reduce asymptotically to the PT-symmetric and asymmetric states without detuning.
The nonlinear effects, such as the long-time self trapping and the spontaneous PT-symmetry breaking, eventually reduce to the situation in \ref{stationary states without detuning}. 
One notes that, in the presence of detuning, the condition of PT symmetry $E_{1}=E_{2}^*$ cannot be satisfied.
However, the PT-symmetry-like states can still exhibit similar signatures, as discussed in \ref{Josephson regime}.
Therefore, the polariton BJJ, as a macroscopic quantum system, can be used to experimentally investigate the PT-symmetry breaking by tuning the center-to-center distance between two pillars, where this phenomenon has been observed only for classical systems that mimic a non-Hermitian Hamiltonian.
\cite{Peng&Nori&Bender:NPhy.10.394(2014)(PT-symmetric_whispering-gallery_microcavities)}

\subsection{Photoluminescence}

\begin{figure*}
  \begin{minipage}{0.49\linewidth}
  \begin{flushleft}
  \textbf{(a)} \\
  \includegraphics[width=\linewidth]{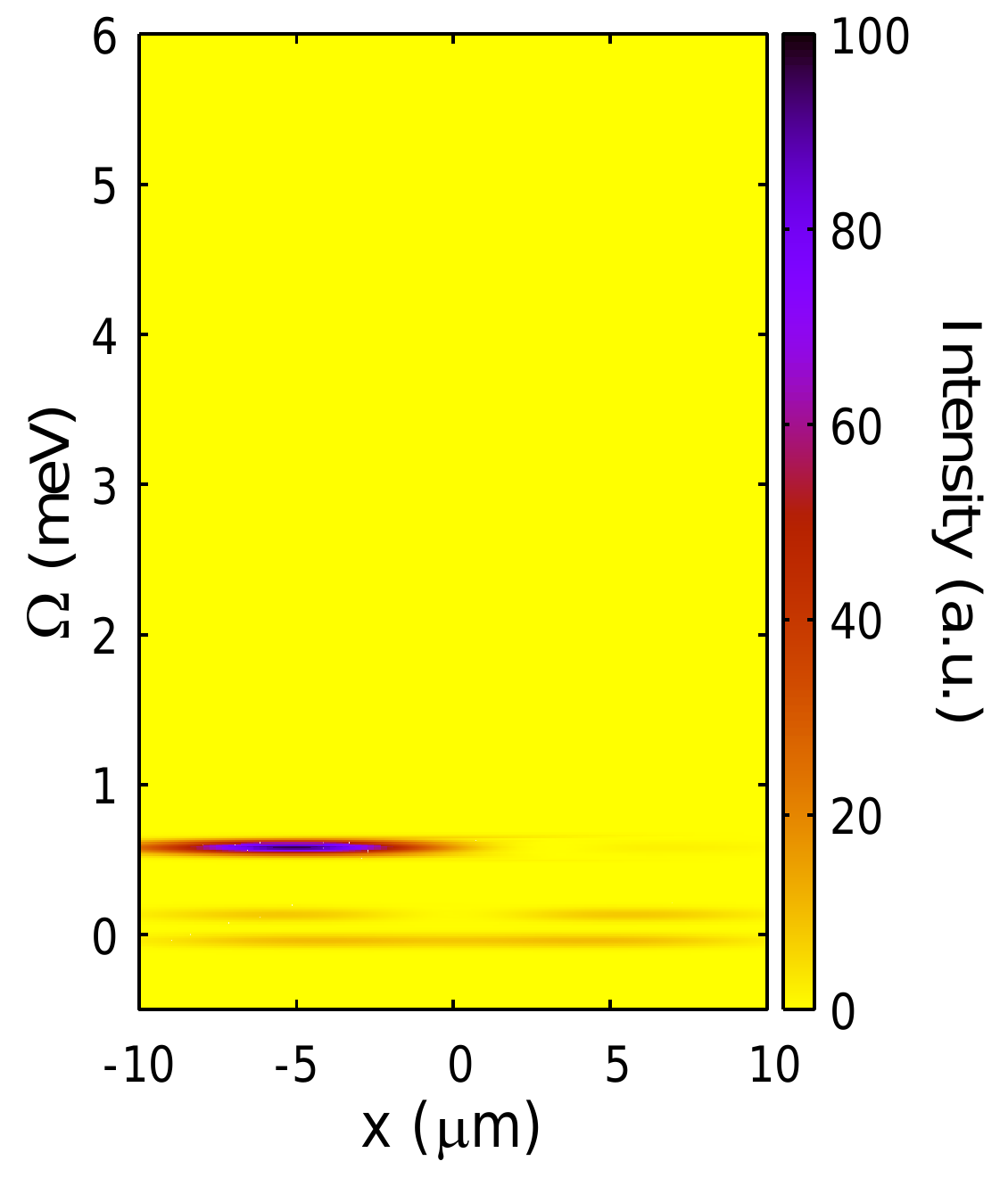}
  \end{flushleft}
  \end{minipage}
  \begin{minipage}{0.49\linewidth}
  \begin{flushleft}
  \textbf{(b)} \\
  \includegraphics[width=\linewidth]{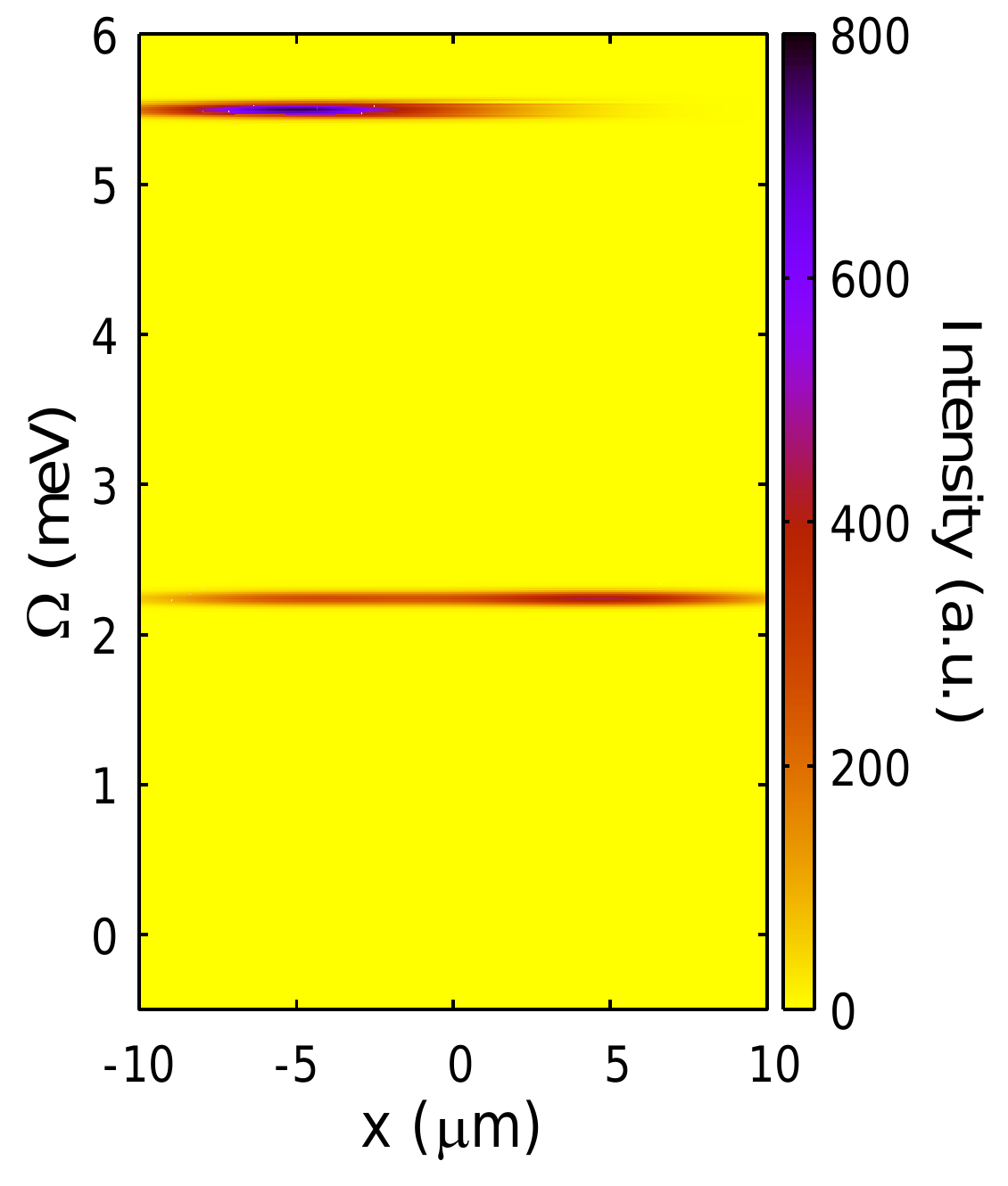}
  \end{flushleft}
  \end{minipage}
  \caption{(Color online)
  Simulation of the spectrally-resolved emission distributions derived from the stable states (a) close to the threshold (with $P_{1}=11$) and (b) at high pumping (with $P_{1}=50$ and an additional potential difference $\Delta\varepsilon_{12}=1$ meV induced by the incoherent polaritons and high-energy excitons).
  The micropillars are centered at $x_{1/2}=\mp5$ $\mu$m with radii 5 $\mu$m.
  The emission strengths and distributions of the $n$-th stable state $|\Psi_{n}(x,\Omega)|^2=|\Psi_{1n}(x-x_{1},\Omega-\Omega_{n})+\Psi_{2n}(x-x_{2},\Omega-\Omega_{n})|^2$ are determined by its energy $\Omega_{n}$, particle numbers, population imbalance, and phase difference, where the spectral and spatial distributions on each pillar $\Psi_{jn}(x-x_{j},\Omega-\Omega_{n})$ are defined by Gaussian functions.
  The three states in (a) from bottom to top are the PT-symmetric $\Omega_{-}$ state, the PT-symmetric $\Omega_{+}$ state, and the self-trapped state, respectively. 
  In (b), the PT-symmetric $\Omega_{+}$ state becomes unstable and only two states are left.
  }\label{Polariton-BJJ PL}
\end{figure*}

Our results can give further insight on the micropillar experiment,
\cite{Galbiati:PRL.108.126403(2012)(Polariton_BEC_Photonic_Molecules)} 
where several states with different energies are found above the threshold in the spectrally-resolved emission distribution. 
Similar phenomena were also observed in CdTe microcavities with spatial photonic potential disorder.
\cite{Love:PRL.101.067404(2008)(Intrinsic_decoherence_coexisting_states_polariton_BEC),Krizhanovskii:PRB.80.045317(2009)(Coexisting_states_disordered_polariton_BECs)}
This phenomenon can be attributed to the coexistence of multiple stable states resulting from incoherent initial conditions or noises.
\cite{Love:PRL.101.067404(2008)(Intrinsic_decoherence_coexisting_states_polariton_BEC),Aleiner:PRB.85.121301(2012)(Radiative_coupling&weak_lasing_polariton_BECs)}
When the initial condition is not specifically controlled by an additional resonant pumping 
as in Ref.~\onlinecite{Abbarchi:NPhy(2013)(Polariton_BEC_Josephson_oscillations)}, 
the fluctuations of population imbalance and phase difference could cause simultaneous condensation into multiple stable states.
This can be analyzed by adding a noise term into the GPE.
\cite{Aleiner:PRB.85.121301(2012)(Radiative_coupling&weak_lasing_polariton_BECs)}

Figure \ref{Polariton-BJJ PL} shows simulations of the spectrally-resolved emission distribution from the stable states in Fig.~\ref{Stationary states & multi-stability}.
Although the interactions 
among these states are ignored in our simplified model, the signatures in the experiment of Ref.~\onlinecite{Galbiati:PRL.108.126403(2012)(Polariton_BEC_Photonic_Molecules)} can still be qualitatively captured by the multi-stability.
Close to the threshold [Fig.~\ref{Polariton-BJJ PL}(a)], there are three available states, i.e., the PT-symmetric states (both bonding and antibonding) at the two bottom levels as well as the self-trapped state at the highest level. 
The self-trapped state has a much stronger condensation than the other two states owing to the localization of the stimulated scattering. 
A spatial shift of both the bottom states to the unpumped site was observed in the experiment, and this can be attributed to the repulsive interactions with the self-trapped states and the pumping spot.

At high pumping [Fig.~\ref{Polariton-BJJ PL}(b)], the anti-bonding symmetric state becomes unstable, and the emission spectrum is dominated by the self-trapped state and the bonding symmetric state. 
As mentioned, a positive detuning must be introduced by the interactions with the incoherent polaritons/excitons, leading to an imbalance of the bottom state weighted to the unpumped site, in agreement with the experimental observation. 
The detuning does not affect the self-trapped state much.
However, the bottom state could be more localized in site-2 due to the interaction with the self-trapped state.
When the pumping strength is larger, the reservoir polaritons are not accumulated more due to the depletion by condensation.
On the other hand, the incoherent excitons which are inactive to the polariton condensation can increase with pumping, and thus the detuning increases.
As the number of condensate polaritons also increases, the detuning in Eq.~(\ref{Josephson eqs. with current source}) can be compensated, with a finite imbalance $\zeta<0$.
As a result, the PT-symmetric-like state exists at large pumping unless Eq.~(\ref{Spontaneous PT-symmetry breaking}) is violated.


For the microcavities with potential disorder,
\cite{Krizhanovskii:PRB.80.045317(2009)(Coexisting_states_disordered_polariton_BECs)}
the observed emission intensity lacks of the time-reversal symmetry, i.e., $I(k)\neq I(-k)$, and the phase differences of the wave functions are locked to $\pi/2$.
Under equivalent pumping, with the time-reversal symmetry kept, a radiative coupling can be introduced to explain this symmetry breaking and the phase locking.
\cite{Aleiner:PRB.85.121301(2012)(Radiative_coupling&weak_lasing_polariton_BECs)}
However, the radiative coupling seems to be unimportant for micropillars.
First, the observed phase difference for self-trapping is locked to $\pi$ instead of $\pi/2$.
\cite{Abbarchi:NPhy(2013)(Polariton_BEC_Josephson_oscillations)}
Second, introducing this coupling to our model, i.e., $J\rightarrow J+i\Gamma$, will essentially break the PT symmetry in Eq.~(\ref{On-site energy}) and lead to the absence of the bottom states in Fig.~\ref{Polariton-BJJ PL}, where only the self-trapped state with a $\pi/2$ phase is left.
Accordingly, the Josephson coupling dominates the micropillar system and the radiative coupling can be ruled out.
In addition, a $\pi/2$ phase is also possible for both the PT-symmetric states near the exceptional point, where $\sin(\Delta\varphi)\approx1$ in Eq.(\ref{Josephson eqs. with current source}).


\section{Summary}

In summary, we have analyzed, within mean-field theory, the dynamics of a polariton Josephson junction pumped on one side. 
The reservoir polaritons contribute to a bias voltage through interacting with the condensate polaritons, as well as a d.c.~current source by stimulated scattering together with the decay rates.
In contrast to the equivalent pumping, there is no long-lived a.c.~Josephson oscillation.
Instead, the Josephson current induces multiple stable states corresponding to different initial conditions.
These states can be attributed to either the self-trapping effect resulting from the nonlinearity or the parity-time symmetry of the system.
The coexisting states with different energies and the $\pi$-phase locking observed in recent experiments can be explained by our results.
We also predict the condensation and a hysteresis phenomenon below the threshold.
Moreover,  the spontaneous parity-time symmetry breaking can be related to the competition between the d.c.~current drive and the critical current of the junction.
As a result, this macroscopic quantum system can be used to investigate such a phenomenon which has been observed only in classical systems.

\acknowledgments

This work is partially supported by the National Center for Theoretical Sciences and National Science Council, Taiwan, Grant No.~NSC 101-2628-M-006-003-MY3 and Grant No.~MOST 103-2112-M-006 -017 -MY4. 
FN is partially supported by the RIKEN iTHES Project, MURI Center for Dynamic Magneto-Optics, and a Grant-in-Aid for Scientific Research (S).

\bibliographystyle{apsrev4-1} 
\bibliography{QPT_aps}


\end{document}